\pdfoutput=1
\documentclass[universe,article,accept,pdftex,moreauthors]{Definitions/mdpi} 
\firstpage{1} 
\pubvolume{12}
\issuenum{7}
\articlenumber{195}
\pubyear{2026}
\copyrightyear{2026}
\externaleditor{~} %
\datereceived{29 April 2026} 
\daterevised{19 June 2026 } %
\dateaccepted{23 June 2026 } 
\datepublished{29 June 2026} 
\Title{Dark Halos, Intermediate-Mass BH and Nuclear Stellar Clusters in Dwarf Spheroidals}

\Author{ {Rojin} %
 Bayati $^{1}$ and Paolo Salucci $^{2}$*}

\address{%
$^{1}$ \quad Department of Physics and INFN,  {University of Genova,} %
 Via Dodecaneso, 33, 16146 Genova, Italy; rojin.bayati@edu.unige.it\\
$^{2}$ \quad  {SISSA,} %
 Via Bonomea 265, 34136 Trieste, Italy}

 \corres{Correspondence: salucci@sissa.it}

\abstract{ 
 We exploit the halo densities of three dwarf spheroidals recently obtained %
   to investigate their dark matter distribution. We address the well-known DM halo core-cusp issue by finding that the NFW $\Lambda $CDM halo profile is in strong disagreement with these inferred halo densities, which instead result %
  very well-reproduced by the Burkert halo cored profile with the central density $\rho_0$ and the core radius $r_0$. 
These structural parameters result %
 connected among themselves and with the length scale of the stellar matter component in a way similar to that present in disk systems. By analyzing the properties of dark halo densities in their innermost regions, we find no trace of a central intermediate-mass black hole or of a nuclear stellar cluster. }

\keyword{dark matter; cosmology; dwarf spheroidals}

\begin{document}

\section{Introduction}

 The nature of dark matter in galaxies, where its presence is certain \citep{R80,B81} (e.g., \citep{PSS}), is still unknown. In disk systems there is very strong evidence that the dark halos that envelop them do not have an inner cuspy density profile as do %
  those that emerge in cosmological N-Body simulations performed of the scenario of collisionless cold $\Lambda$CDM (i.e., the NFW profile \citep{NFW}). They in fact show a wide inner region in which the density of the dark halo is almost constant (see Sec. 8.2 in the review \citep{Salucci2019_DM_galaxies} and references therein, see also \citep{Gentile2004,dba, db2001, KF,deBlok2010_core_cusp}). 
Let us stress that the phenomenon of the ``cored'' DM halo density distribution generally involves one of the following specific halo profiles that, alongside a baryonic component, is found to reproduce the disk kinematics well: (a) the Pseudo-Isothermal, $\rho_h(r)=\rho_0/(1 +r^2/r_c^2) $ (\citep{Salucci2019_DM_galaxies}), (b) the halo profile in the Universal Rotation Curve \citep{PSS}, $\rho_h(r)=\rho_0 \ \frac{a^2 \left(3 a^2+r^2\right)}{3 \left(a^2+r^2\right)^2} $, and (c) the Burkert profile, see Equation~\eqref{eq:BUR} (\citep{BU,S07}). The first two profiles, however, do not converge to the NFW profile when $r>1/3\ R_\mathrm{vir}$, a constraint that seems required due to weak lensing measurements (e.g., \cite{EXT,Gavazzi}) and from a careful extrapolation of the DM halo densities derived at inner radii \citep{S07}. Therefore, the Burkert (hereafter, B) profile emerges, in disk galaxies, as the reference ``cored'' halo distribution: specifically, no other cored profile has been equally successfully tested to achieve such a characterization \citep{Salucci2019_DM_galaxies}.  %
 
It is also important to notice that the issue of the cored nature of the DM halos around disk systems (\cite{Salucci2019_DM_galaxies}) and, specifically, around dwarf irregulars (dIrrs) \citep{K17}, disk systems that better correspond to the dSphs, does not consist only of the well-known disagreement between the halo density profiles derived from the galaxies' Rotation Curves (RCs) and the NFW ones emerging from cosmological simulations performed on a $\Lambda$CDM scenario. 
In fact, the structural parameters ( {{\bf SP}s}%
) of the DM halos, i.e., the central density $\rho_0$ and the core radius $r_0$, are found to correlate among themselves and with the stellar disk length scale $R_D$, a fact not even conceivable in the above scenario
 \citep{Donato2004_cores_scalelengths,S07, Donato2009_const_surface_density}, see also Sec. 10 of the review \citep{Salucci2019_DM_galaxies}.
 In the collisionless particle scenario, in fact, it is very difficult to envisage how the detected correlations between  {{\bf SP}}s of SM particle components and those of the dark particle component could have arisen. This evidence must be taken into serious consideration in that it reveals a complex interplay between dark matter and luminous matter, and it may indicate the existence of a dark particle that, unlike those considered so far, interacts with itself and/or with standard model particles.\endnote{As this work is concerned with dwarf galaxies, which are totally DM-dominated, we do not consider the possibility that the current structure of the dark halos was created by some baryonic mechanism (e.g., Supernovae feedback, \citep{nunez, Pontzen}).} 

In dSphs, the Hubble type under investigation here, the stellar density distribution is well-reproduced by a Plummer sphere with (projected) half-light radius $R_h$ \citep{Plummer}:
 \begin{equation}
\rho_P(r) = \frac{3M}{4\pi R_h^{3}}
\left(1 + \frac{r^{2}}{R_h^{2}}\right)^{-5/2}
\end{equation}
 {Differences/similarities} %
 in the dSph DM structure with respect to that in dwarf disks (dIrr) of similar virial mass $M_ \mathrm{vir}$ can provide us with valuable information on the nature of the dark particle. 

It is useful to compare $R_h$ with $R_D$, the standard length scale of the stellar disks with Freeman surface luminosity/mass $\propto \exp \, (-\frac{R}{R_D})$ \citep{Freeman}. In the latter systems, the radius $R=R_D$ marks the region containing 24.5\% of the total stellar mass; in a Plummer spheroid, the same mass fraction occurs at the radius $r=0.833 \; R_h $. Therefore, we can define
\begin{equation}
R_D^{sph}\equiv 0.833\, R_h
\label{eq:RDPLUMMER}
\end{equation}
as the radius that, in spheroids, corresponds to the quantity $R_D$ in disk systems (reported for the three objects in  {Table}%
~\ref{tab:inner_radii}).

Let us remind the readers of the reasons for the interest in the DM distribution of dwarf spheroidal (dSph) galaxies. They are among the smallest objects in the universe and, unlike disk systems, are entirely dominated by the dark matter component, even in their innermost regions \citep{Walker2009_universal_mass_profile, Walker2012_dSph_review, Salucci2012_dSph_scaling,Gold,LeoI,ANDRADE}. Owing to this extreme dark matter dominance, dSph halos are expected to be only weakly affected by baryonic processes, such as the adiabatic contraction induced by baryonic infall or the core-forming feedback from supernova explosions. Their virial masses have a range of about two orders of magnitude, allowing the investigation of DM structural properties as a function of the halo mass $M_\mathrm{vir}$ at the lowest values in galaxies of this~quantity. 

The dSph kinematics consist of an ensemble of independent measures, each of them coming from a single star. Until recently, the advantage that dSphs do not require the complex dark-luminous mass decomposition needed in disk systems has been paired with the fact that, in these systems, the 3D kinematics and distribution of the tracers of the gravitational field are not fully known. As a matter of fact, to mass model these galaxies, one has to resort to the complex Jeans Equations, rather than to the much simpler Poisson Equation used in disk systems. In addition, until recently, dSphs had a limited number of stars whose motions were measured \citep{Battaglia2013_dSph_Jeans}. All this triggers large statistical errors and large systematic uncertainties, making it difficult to probe the distribution of DM in these systems (see the Introduction of \citep{yang2025darkmattercontentmilky} and references therein).

New avenues for investigating such mass distribution have come up with the work by \citep{yang2025darkmattercontentmilky} (hereafter Y+25) in which they exploited very recent Milky Way Survey observations, performed by the Dark Energy Spectroscopic Instrument (DESI), %
 of three classical dSphs: Draco, Sextans and Ursa Minor. We briefly report relevant aspects of their careful analysis, which has provided us with the DM halo density of these dSphs. As a first step, they investigated the membership of each candidate star. To minimize the number of interlopers, they selected as tracing stars only those with a high probability ($\geq$0.9) of being members of these systems that were found to lie within $\sim$2.5 half-light radii $R_h$ from the dSph centers. The DESI sample includes 96 stars for Draco, 240 for Sextans, and 1003 for Ursa Minor, to which they added previous data from \citep{Walker2023_dSphSurvey} containing 374, 368, and 416 stars for Draco, Sextans, and Ursa Minor, respectively. Noticeably, the kinematical data \citep{yang2025darkmattercontentmilky} extend out to about 3 kpc in each galaxy, a radius whose value corresponds to (9--18) times that of $R_D^{sph}$. This and the large number of available star tracers have provided one with kinematics with very good radial coverage of an unprecedented distance from the galaxy centers; this allows a robust determination of the mass distribution of the three objects. 
 
For the three dSphs, \citep{yang2025darkmattercontentmilky} determined the DM density profiles $ \rho(r)$ and their uncertainties\endnote{Data are available at %
 \url{https://zenodo.org/records/16919342}, accessed on 3 September 2025.} from the observed line-of-sight velocities of the tracer stars by performing on them a Jeans Anisotropic Multi-Gaussian Expansion modeling (JAM) \citep{Watkins2013,Emsellem,MGE}.
 The details of the process through which, without resorting to an a priori specific profile, the DM halo density $\rho(r) $ was inferred from the kinematics of about 1 k tracer stars per galaxy are discussed in detail in Yang+25. Let us notice that this refined dynamical analysis adopted the framework of the {\it  {empirical}%
} gNFW \citep{Zhao_1996} dark halo density profile, a particular case of the Zhao profile \citep{Zhao_1996}:
\begin{equation}
\rho_\text{gNFW}(r)=\frac{\tilde\rho_{s}}{\big(\frac{r}{\tilde r_s}\big)^{\gamma}\big(1+\big(\frac{r}{\tilde r_s}\big)\big)^{3-\gamma}}\,
\label{eq:Zhao}
\end{equation}
where $ \gamma$ is the inner slope of the profile and $ \tilde \rho_{s}\,, \tilde r_s$ are the reference radius and the reference halo density. Adopting this {\it  {empirical}} profile for the halos of the three dSphs is very suitable. In fact, for dSph, a clear support/disproof for the B or the NFW profiles is presently lacking, unlike in the case of disk systems. It is then correct and useful to adopt a {\it  {neutral}} profile that, via its additional free parameter, can account for cored, cusped and in-between DM halo profiles\endnote{In particular, despite its name, this profile can account for distributions that are in complete disagreement (both numerical and physical) with the NFW one.}. Noticeably, this strategy is not jeopardized by the presence of a gravitationally relevant stellar component, as in spirals, or by a low number of kinematical measurements, which is %
 insufficient to statistically damp their random errors as in old dSph investigations.

 Recently, in dwarf irregulars (dIrrs), as in other disk systems, relevant properties of their DM halos, including their cored nature, have emerged from analyzing their halo density profiles{\citep{K17}}. Here, we accurately investigate the density profiles of three dwarf spheroidals (dSphs) obtained by Yang+25, and perform on them the same investigation devised for (dwarf) disk systems. The derivation of the above density profiles relies, as it generally does %
  in dSphs, on several assumptions whose consequences on the inferred densities are briefly discussed here. Let us remind the reader that the above work employs the Axisymmetric Jeans Anisotropic Multi-Gaussian Expansion method to (mass) model the available kinematical data, a method that is generally considered very efficient and robust (for details, see  {Section 3.1} %
 of Yang+25). It is proper to stress that the large number of member and tracer stars that are available for the three dSphs makes the MGE aspect of the JAM method very robust. Moreover, one has to point out that, as usual in this kind of analysis, Yang+25 have assumed that the velocity and the stellar density ellipsoids are aligned and that the tracer's velocity anisotropy is constant. As regards the crucial and, in principle, difficult choice of the DM profile to adopt in JAM, let us remind the reader that, in dSphs, in contrast to what occurs in spirals and in ellipticals, the stars that trace the total gravitational field kinematically are subject to the distribution of (only) the dark component. No tricky dark-luminous mass decomposition is required to resolve the galaxy kinematics in its dark and luminous components: these objects are just spheres of dark matter. As a result, the assumed gNFW profile, considered here as an empirical and model-independent profile, is flexible enough to represent {\it  {both and at the same time}} the B and the NFW halo profiles. This strategy leaves to the data itself the task of deciding which one is correct.\endnote{Noticeably, in the case of a non-negligible contribution of the stellar component to the galaxy gravitational potential, this approach will not work in that, in reproducing the observed kinematics, the values of the fraction of dark-to-stellar mass and of the halo structural parameters become very degenerate among themselves, and the gNFW halo does not provide a short-cut to solving the cusp-core issue.} 
 
 Let us remark that it is not within the aims of the present work to go deeper than Yang+25 in the derivation of the halo densities in these galaxies, but instead to complete the investigation of the inferred density profiles with respect to certain important physical aspects.

 Finally, the gNFW profile is useful for tracing deep %
  inner spherical components distributed on a scale much smaller than the B/NFW characteristic scales $r_0$ or $\tilde r_s$, namely, a central intermediate-mass BH or a nuclear stellar cluster for which, in this region, $M_\text{gNFW}(r)=M_\text{IMBH,NSC}(r)$.

\begin{figure}[H]
 \includegraphics[width=0.48\textwidth]{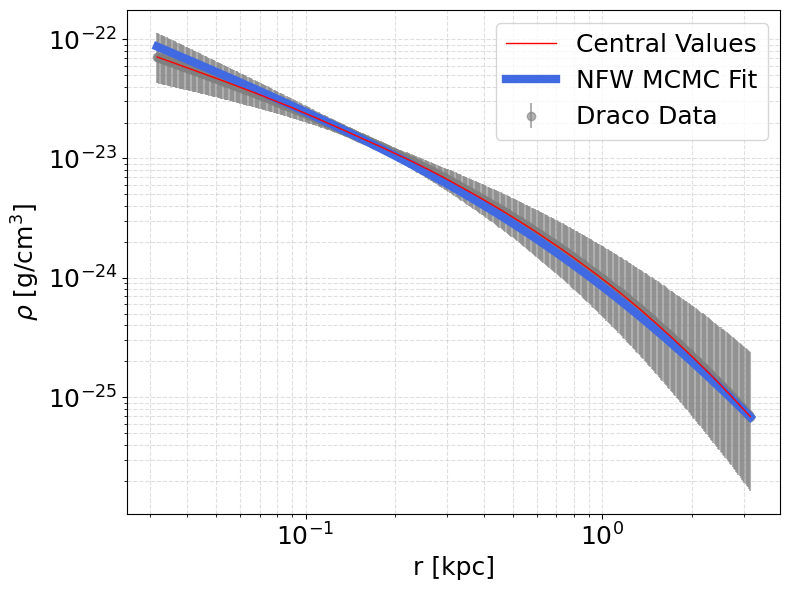}
 \includegraphics[width=0.48\textwidth]{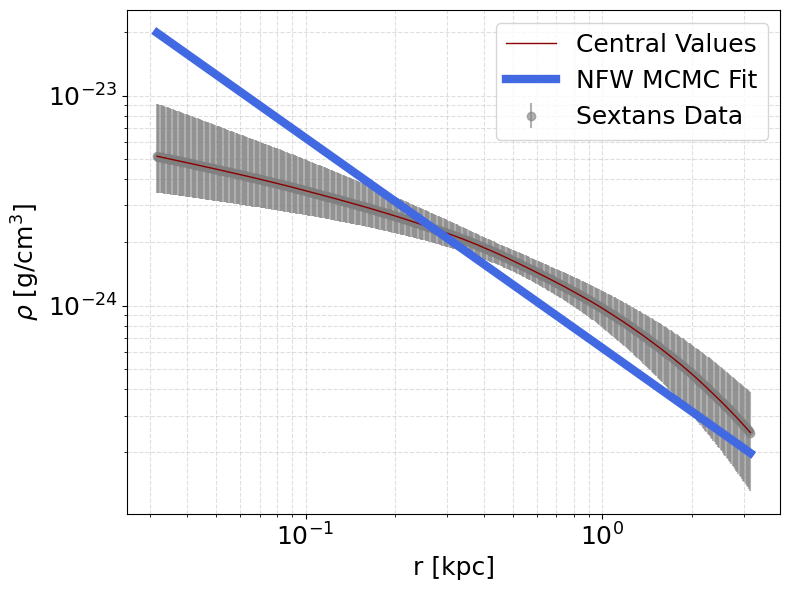}
 \includegraphics[width=0.48\textwidth]{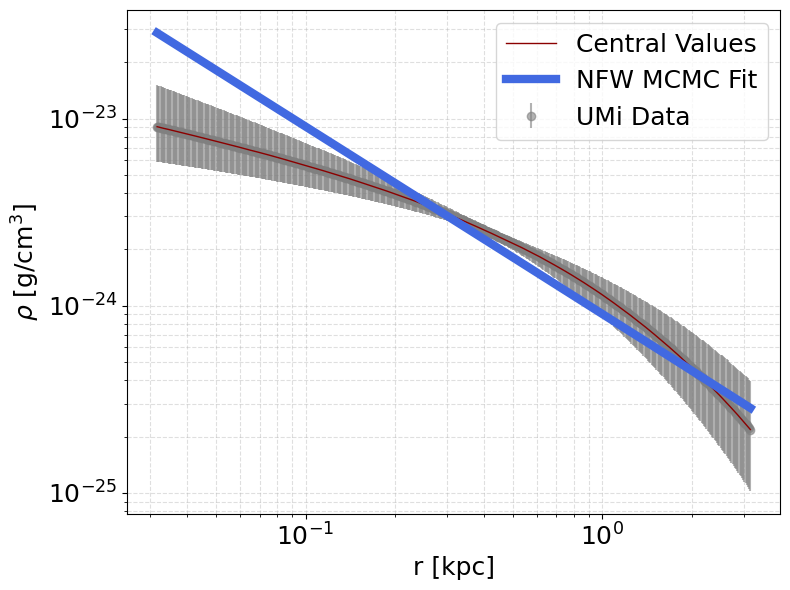}
 \caption{ {Dark} %
 matter density profiles of Draco, Sextans, and UMi compared with their best-fitting NFW models. Gray points denote the inferred densities with uncertainties, and the blue curves show the corresponding NFW fits. The red curves connect the central values of the measurements.}
 \label{fig:NFWfits}
\end{figure}

 As a result, for each galaxy, the inferred halo quantities\endnote{Available following Footnote 1.} can be compacted in three \mbox{3 $\times$ 500} matrices: $A_n (r_k,\rho(r_k), \delta \rho(r_k))$, where $n$ runs from 1 to 3 tagging the galaxies, $k$ runs from 1 to 500, and $\rho(r_k)$ is the inferred density at a radius $r_k$, uniformly located inside the region limited by the innermost radius $R_i=3.2\times10^{-2}$~kpc and the outermost radius $R_o=3.2$~kpc, for which kinematical data are available. $\delta \rho(r_k)$ is the 1-$\sigma$ uncertainty of $\rho(r_k)$~\citep{yang2025darkmattercontentmilky}. For each object, the latter is shown as points with their uncertainties as error bars in Figure~\ref{fig:NFWfits}. Let us note that the uncertainties, which are as big as $\sim$(0.2--0.25)  {dex}%
, are not negligible, so one should consider the inferred densities + their uncertainties as an area in  {the} %
 $log\, \rho - log\, r$ plane, inside which any reasonable model density profile $\rho_{mod}(r)$ should lie.

The results of \citep{yang2025darkmattercontentmilky}, namely the inferred DM halo densities, shed light on the core/cusp issue. Using the inner slope $\gamma$ of the adopted empirical gNFW profile as the indicator of the cuspiness of the dSph halos, the resulting values $0.26 \leq \gamma \leq 0.71$ led them to claim evidence against a cuspy DM distribution in at least two objects. On the other hand, the same considerations indicate that these halos are not maximally cored, i.e., they do not exhibit constant density out to $R_h$.

The data in the matrices $A_n $ allow us to investigate the dark dSph halos from other effective points of view. The kinematics are available from $r=R_i$ to $r=R_o$, i.e., over a wide region where the NFW and the cored halo are expected to have markedly different profiles. In spiral galaxies and in dwarf irregulars, the failure of the NFW profile is evident, not only at $r \rightarrow 0$, but also at all radii $r$, with $r \leq \ 3\ R_D\sim R_o/2$  {(}%
 see: \citep{Salucci2019_DM_galaxies}). Furthermore, failure does not always emerge as the inability of the NFW halo, along with the stellar disk component, to successfully fit the circular velocity $ V(r)$. In several cases, the fit is successful, but the resulting NFW halo parameters do not comply with the concentration vs. virial mass relation that emerges in $\Lambda$CDM N-Body simulated halos \citep{Dehghani}. 

Therefore, the cored/cuspy situation must be resolved in a much more complex way than assessing the values of the DM halo density in the innermost regions of galaxies. We take advantage of the work of \cite{yang2025darkmattercontentmilky} that, for three dSphs, has provided us with data, collected in the matrices $A_n$, suitable to investigate such complexity. First, we check whether the NFW density profile fails to reproduce the inferred halo density in the ways discussed above for the disk systems, including the dIrrs (\citep{K17}), i.e., disk galaxies with luminosities and virial masses similar to those of the present dSphs. Furthermore, in dIrrs and in other Hubble-type objects, the cored DM halos are found to obey particular relations entangling the dark halo central densities, the core radii and the stellar disk length scales. This scenario is a decisive signature of great physical importance of the cored aspect of the DM halos, and it serves as a portal to the nature of the dark particle. Again, matrices $A_n$, along with the galaxies' photometric measurements, check the three dSphs for the existence of the above relationships. Let us remind the reader that this work aims to start from Yang+25 and to complete the rendition of their results within the B and the NFW scenarios.

Moreover, in these systems, the availability of measurements of the DM densities inside a very central region extending from $30$ pc to 100  {pc} %
 and the lack there of a significant number of stars from the stellar spheroids allow us to investigate the presence of a central intermediate-mass black hole and that of a nuclear star cluster that, if existing in this region, are likely to overcome the DM halo mass.

The structure of this paper is as follows: in Section \ref{sec2} we analyze the general NFW density profile and present the corresponding fits to the dSphs data, and in Section \ref{sec3} we repeat the analysis for the B profiles and compare their relative performance. Section \ref{sec4} examines the overall structure of dSph halos by highlighting specific empirical relations shared with dwarf irregulars. In Section \ref{sec5} we focus on the innermost regions of dSphs, discussing whether they host an intermediate-mass black hole or a nuclear stellar cluster. Section \ref{sec6} summarizes our main conclusions.

 \section{NFW DM Halos}\label{sec2}

 In the $\Lambda$CDM scenario, the DM halo density, the result of N-Body cosmological simulations, is well-reproduced by the NFW profile \citep{NFW}: 
\begin{equation}
\rho_{\text{NFW}}(r; r_s,\rho_s)=\frac{\rho_{s}}{r/ r_s\big(1+ (r/r_s)^2\big)} 
\label{eq:NFWrho}
\end{equation}
 {The} corresponding enclosed mass, at a radius $r$, is 
\begin{equation}
M_\text{NFW}(r) = 4\pi \rho_s r_s^3 \left[ \ln\left(1+\frac{r}{r_s}\right)+ \frac{r}{r+r_s} \right].
\label{eq:nfw_mass}
\end{equation}

The virial mass of a halo is defined as the mass enclosed within the virial radius 
 $R_{\mathrm{vir}}$, the radius at which the average density of the halo 
is a fixed multiple $\Delta_c$ of the critical density of the universe: \vspace{6pt}

\begin{equation}
M_{\mathrm{vir}} = \frac{4}{3}\pi R_{\mathrm{vir}}^3 \, \Delta_c \, \rho_{\mathrm{crit}},
\label{eq:mvirdef}
\end{equation}
with the critical density given by $\rho_{\mathrm{crit}} = \frac{3H_0^2}{8\pi G}$. The virial radius is then determined by requiring that the enclosed mass of the NFW profile matches the cosmological definition of the virial mass\endnote{In this work: $\Delta_c =180$, $\rho_{\mathrm{crit}}=1 \times 10^{-29}$ g/cm$^3$. Results do not change by assuming other reasonable values.} $M(R_{\mathrm{vir}}) = \frac{4}{3}\pi R_{\mathrm\mathrm{vir}}^3 \, \Delta_c \, \rho_{\mathrm{crit}}$.
 This profile has two  {{\bf SP}}s for each halo: the reference density $\rho_s$ and scale radius $r_s$. The halo structure also emerges from the relationship ($c\equiv R_\mathrm{vir}/r_s$):
\begin{equation}
 c(M_\mathrm{vir})= 9.35 \Big(\frac {M_\mathrm{vir}} {10^{12}\, M_\odot} \Big)^{-0.13} 
 \label{eq:cmvir}
\end{equation}
e.g., \citep{Bullock}, which %
 connects the above two parameters.

Let us stress that the asymptotically cuspy behavior of the $\Lambda$CDM DM halo density $\rho_{\text{NFW}}(r)$ is (mathematically) described by the value of the slope: 
\begin{equation}
 \Gamma_{\text{NFW}}(r)= \lim_{r \to 0} \frac{d \log \,\rho_{\text{NFW}}(r)}{d \log\, r}=-1\end{equation} 
 that results in agreement with the observed slope 

\begin{equation}
 \Gamma(r)= \lim_{r \to 0} \frac{d \log \,\rho(r)}{d \log\, r}\end{equation} 
 
However, in real galaxy measurements, as in the case of the present dSphs, even for the innermost kinematical measure at the radius $R_i$ we have that $R_i>0$; then, considering that in the region around $r \simeq 0$, $\Gamma_{\mathrm{NFW}}(r)$ declines very rapidly with radius, we have that the {\it  {expected}}  {value} %
 $\Gamma_\mathrm{NFW} (R_i)$ can be significantly lower than $-1$ and down to $-$1.4, depending on the value of $R_i/r_s$ \citep{Salucci2025}. Measuring $\Gamma(R_i) =-1$ does not guarantee an NFW halo density. 

For each dSph, we fit the 500 available inferred densities $\rho(r_k)$\endnote{See Footnote~2.} with a pure NFW model characterized by two free SPs: the scale density, $\rho_s$, and the scale radius, $r_s$\endnote{The code used in this work is available at %
 \href{https://github.com/rojinbayati/dm-mcmc-fits}{https://github.com/rojinbayati/dm-mcmc-fits}, accessed on 27 June 2026}. Parameter estimation was performed within a Bayesian framework adopting uniform priors of $10^3 < \rho_s < 10^{10}\,\mathrm{M_\odot\,kpc^{-3}}$ and $0.5 < r_s < 250\,\mathrm{kpc}$. The posterior distributions were sampled using the affine-invariant ensemble Markov Chain Monte Carlo algorithm implemented in the \texttt{ {emcee}%
} package \citep{Goodman2010, ForemanMackey2013}. Convergence of the Markov chains was verified before deriving the final parameter estimates. The reported parameter values correspond to the medians of the marginalized posterior distributions, while the uncertainties represent the 16th and 84th percentiles (i.e., approximately $1\sigma$ credible intervals).

\begin{figure}[H]
 \includegraphics[width=0.7\textwidth]{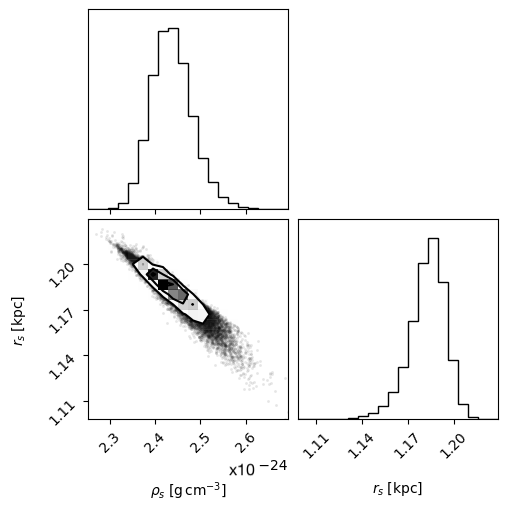}
 \caption{
  {Draco} %
 posterior distributions of the NFW fit for 
 $\rho_s$ and $r_s$ with their 1--3$\sigma$ credible contours 
 and median values  }
 \label{fig:NFWfits-draco-corner}
\end{figure}

 These best-fit parameters were used to construct smooth NFW density profiles, which were compared with the data to assess the quality of the fit. For the NFW scenario, values of the SPs are given in Table~\ref{tab:draco_nfw_params}. The outcome of the fit is evident in Figures~\ref{fig:NFWfits} and~\ref{fig:NFWfitsresiduals} and in the values of Tables~\ref{tab:draco_nfw_params} and~\ref{tab:nfw_diagnostics}, all showing different aspects of the failure of this profile to represent the dSph's halo densities of Sextans and Umi. In detail, for Sextans and Umi the NFW halo fits fail at almost any radius. Furthermore, the best-fitting values of the  {{\bf SP}}s, $\rho_s$ and $r_s$ are too large to have a physical significance (see ~Table \ref{tab:draco_nfw_params}: $r_s \simeq 10 \, \text{kpc}$  {and} %
 $M_\mathrm{vir} \sim 10^{10} M_\odot$). %
  In Draco, the NFW halo fit is excellent, see Figures~\ref{fig:NFWfits}, \ref{fig:NFWfits-draco-corner} %
 and \ref{fig:NFWfitsresiduals}. However, also for this galaxy, the best-fitting  {{\bf SP}}s, reported in Table~ \ref{tab:draco_nfw_params}, take nonphysical values that are in great disagreement with Equation~\eqref{eq:cmvir} and that imply $M_\mathrm{vir}> 2 \times 10^9 \, M_\odot $, i.e., a value $>50 $ times any realistic expectation.

\begin{table}[H]
\caption{ {Median} %
 values of the NFW parameters and the corresponding virial mass for Draco, Sextans, and UMi dwarfs. Uncertainties correspond to $1\sigma$ credible intervals.}
\label{tab:draco_nfw_params}
\begin{tabularx}{\textwidth}{LCCC}
\toprule
\textbf{Parameter} & \textbf{Draco} & \textbf{Sextans} & \textbf{UMi} \\
\midrule
$\rho_{\mathrm{s}}$ (g\,cm$^{-3}$) 
& $(2.43^{+0.05}_{-0.04})\times10^{-24}$ 
& $(6.38^{+0.16}_{-0.15})\times10^{-26}$ 
& $(9.78^{+0.14}_{-0.13})\times10^{-26}$ \\
$r_{\mathrm{s}}$ (kpc) 
& $1.18^{+0.01}_{-0.01}$ 
& $9.95^{+0.04}_{-0.10}$ 
& $9.53^{+0.04}_{-0.08}$ \\
$M_{\mathrm{vir}}$ ($M_\odot$) 
& $(1.58^{+0.01}_{-0.02})\times10^{9}$ 
& $(3.28^{+0.11}_{-0.12})\times10^{10}$ 
& $(5.56^{+0.13}_{-0.20})\times10^{10}$ \\
\bottomrule
\end{tabularx}
\end{table}
\vspace{-10pt}

 We introduce an additional new test to check the ability of the NFW profile in reproducing the observed halo densities of the three dSphs under investigation. One important property of the NFW halos, emerging from the cosmological simulations performed in the $\Lambda$CDM scenario, is the existence of a relationship that, at any radius $r$, links the following quantities: the radius $r$, the (dark) halo {\it  {density}} $\rho_{\text{NFW}}(r)$ at this radius and the corresponding {\it  {mass}} inside $r$, $M_{\text{NFW}}(r)$. These three quantities are %
  connected, independently of the galaxy halo concentration and virial mass, via the relationship (\citep{Gentile07}) (mass in $M_\odot$, density in $\mathrm{g,cm^{-3}}$, and radius in kpc):
 \begin{equation} 
 \log\, M_{\text{NFW}}(r)= \frac{ 3}{4 }\, \log\, \rho_{\text{NFW}}(r)+ \frac{11}{4} \, \log\, r + 26.17 
 \label{eq:logmnfw}
\end{equation}
 
\begin{figure}[H]
\includegraphics[width=0.7\textwidth]{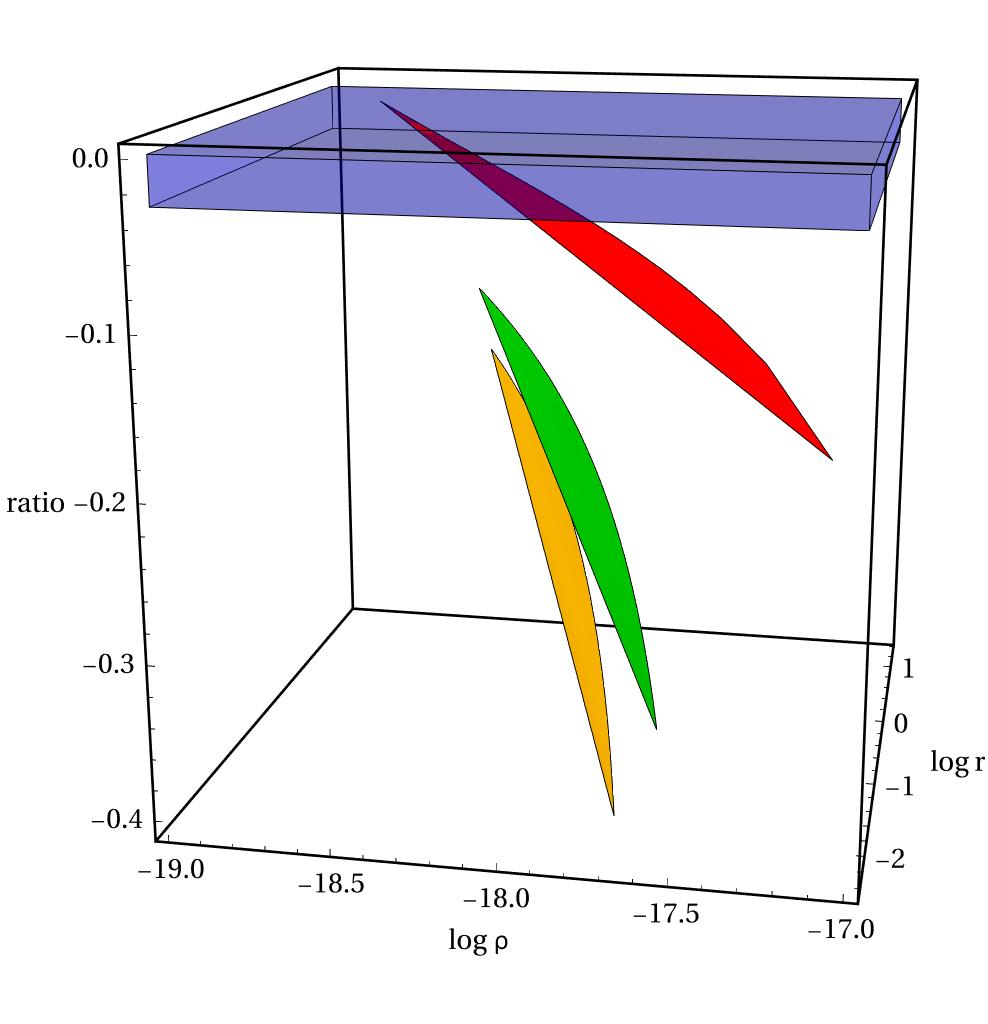}\vspace{-6pt}
 \caption{ {The} %
 ratio $\log\, (M_\mathrm{pNFW} (r)/M_\mathrm{NFW} (r))$ as function of $\rho_\text{NFW}(r$) and $r$ for the three dSph halos. From top to bottom: Draco,Sextans Umi. The blue parallepiped %
  is the region that, in the NFW scenario, should fully contain the three surfaces.}
 \label{fig:Mrhor}
\end{figure}

 We find from simulations that this relation has a very small r.m.s. of 0.05 dex: its emergence in galaxies is a strong signature of an NFW halo. 
 Its physical origin, as that of the NFW profile itself, is not completely understood; however, we can take advantage of the existence, in the NFW framework, of the robust $M(r)-\rho(r)-r$ %
  relationship to break the mass-density degeneracy often present in the mass modeling of galaxies.\endnote{In fact, a B profile is not too different from an NFW profile if one allows large values for the  {{\bf SP}} $r_s$: $r_s> 2 \, r_0$.} We notice that, in objects completely dominated by the dark component, the existence of Equation~\eqref{eq:logmnfw} can be directly verified by means of the galaxy kinematics, without performing the full mass modeling: Equation~\eqref{eq:logmnfw} allows us to investigate the dSph halos from a new angle. Consequently, we adopt the framework of the NFW scenario, so that, at any radius $r_k$\endnote{We use the discrete radial coordinate $r_k$ instead of the continuous coordinate $r$ in that the data are given in such a form, with $k=500$ for each object}:
  \begin{equation}
 \rho_\mathrm{NFW}(r_k)=\rho(r_k)
 \label{eq:rhonfw}
 \end{equation}
 i.e., we force the inferred halo density to follow an NFW profile. Equation~(\ref{eq:rhonfw}) leads to a predicted NFW mass profile given by 
 $$M_\text{pNFW}(r_k)=4 \pi\int \rho(r_k) r_k dr_k$$
  {In} the NFW scenario, this mass inside $r_k$, must be in agreement with $M_{\text{NFW}}(r_k)$, obtained from the RHS of Equation~\eqref{eq:logmnfw}, once combined with Equation~\eqref{eq:rhonfw}. Instead, in Figure~\ref{fig:Mrhor} we realize that, for the three dSphs, the values of the  {ratio} %
 $\log \ (M_\mathrm{pNFW}(r)/M_\mathrm{NFW}(r)$) as a function of radius and local halo density, deviate significantly from zero at all radii, contrary to the expectations of the NFW scenario.

Thus, several independent and converging pieces of evidence imply that the inferred density distribution $\rho(r_k)$ of the three dSphs DM halos cannot be reproduced by an NFW density profile. Our outcome is in substantial agreement with \citep{yang2025darkmattercontentmilky}; however, the direct comparison performed here helps to better clarify the large discrepancy between the $\Lambda$CDM cosmologically simulated halos and those detected around these galaxies. 

The above results, furthermore, make evident how the success of a $gNFW$ halo density profile in best fitting some galaxy kinematical data does not imply that such a profile is (i) a ``quasi-NFW'' one, i.e., of a density profile that might converge to an NFW one as more data become available, or (ii) an actual NFW halo density profile with relevant random measurement errors. The $gNFW$ density profile can be at an irresolvable discrepancy with the NFW one.\endnote{Therefore, the habit of renaming the ($\alpha$, $\gamma$, $-3$) Zhao profile as the $gNFW$ profile is not well-justified.}

\vspace{-12pt}

\section{Burkert DM Halos}\label{sec3}

To reproduce the kinematics of disk systems, the family of the cored halo profiles has a clear champion, the Burkert (B) profile, whose density is written as \citep{BU} 
\begin{equation}
\rho_B(r; r_0;\rho_0)=\frac{\rho_{0}}{\Big(1+ (r/ r_0)\Big)\Big(1+ (r/r_0)^2\Big)} 
\label{eq:BUR}
\end{equation}
 {This} profile has two  {{\bf SP}} running among galaxies: the central density $\rho_0$ and the radius of the ``constant density region'' $r_0$. 
 The corresponding mass, enclosed within a radius $r$, is
\begin{align}
M_B(r)
&= 4\pi \int_{0}^{r} \rho_B(r')\, r'^2\, \mathrm{d}r' \nonumber\\
&= 2\pi \rho_0 r_0^3 \left[
\ln\!\left(1+\frac{r}{r_0}\right)
+ \frac{1}{2}\ln\!\left(1+\frac{r^2}{r_0^2}\right)
- \arctan\!\left(\frac{r}{r_0}\right)
\right].
\label{eq:Mr}
\end{align}

The virial radius $R_{\rm vir}$ is determined, as usual, by solving $\frac{M(R_{\rm vir})}{\frac{4}{3}\pi R_{\rm vir}^3} = \Delta_c \, \rho_c$ and then $M_{\rm vir} = M(R_{\rm vir})$.

\begin{figure}[H]
 \includegraphics[width=0.49\textwidth]{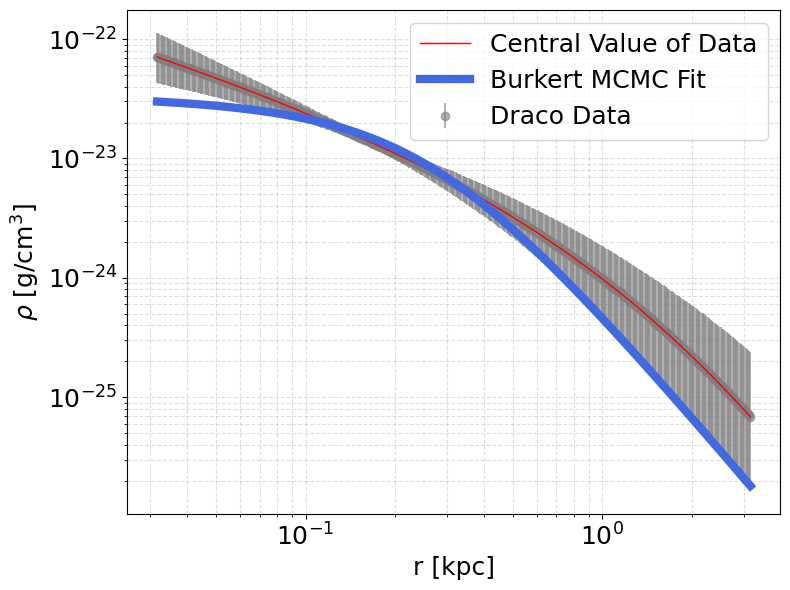}
 \includegraphics[width=0.49\textwidth]{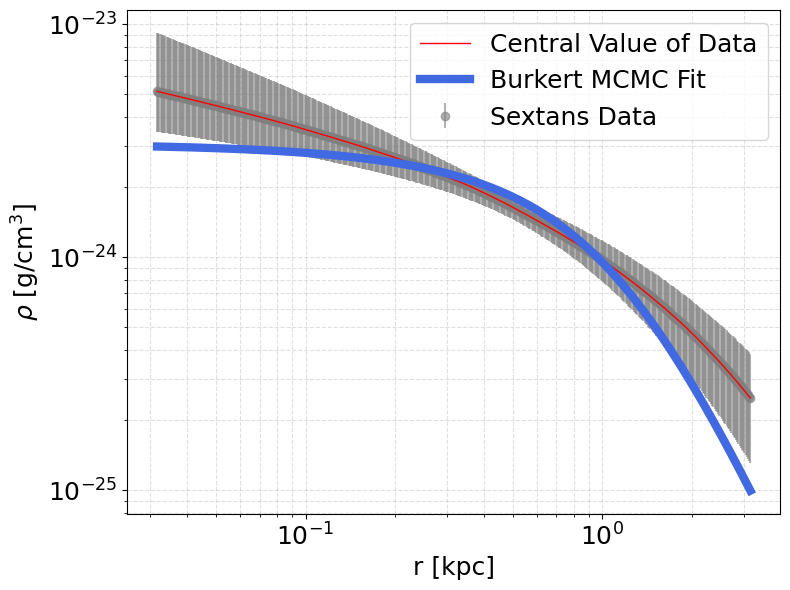}
 \includegraphics[width=0.49\textwidth]{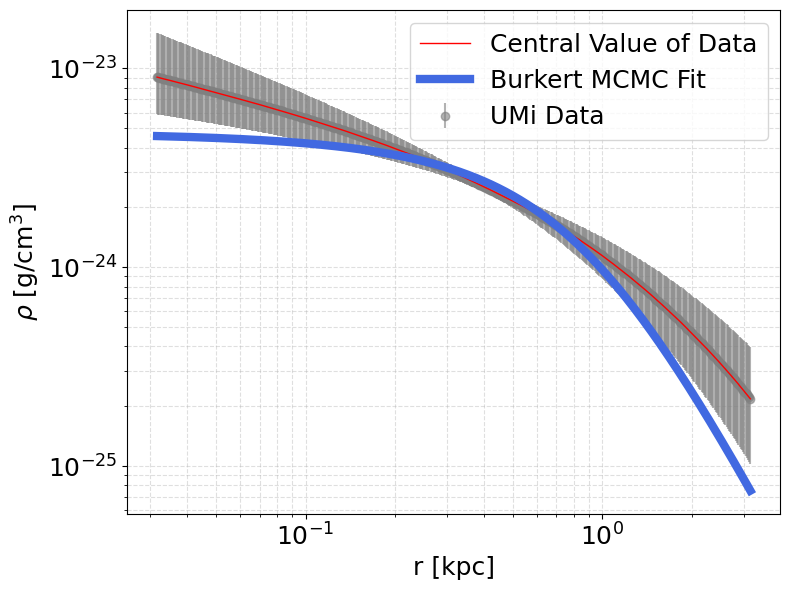}
 \caption{Dark matter density profiles of Draco, Sextans, and UMi compared with their best-fitting Burkert models. Gray points denote the inferred densities with uncertainties, and the blue curves show the corresponding fits. The red curves connect the central values of the measurements.}
 \label{fig:Burkertfits}
\end{figure}

 The results of the best fitting of the inferred densities $\rho(r_k)$ of the matrices $A_n$, performed in the framework of the B scenario by means of the MCMC method described in the previous section, are shown in Figure~\ref{fig:Burkertfits}. The values of the corresponding  {{\bf SP}}s are reported in Table~\ref{tab:burkert_parameters}. We realize that the B halo profile $\rho_B(r)$ provides very good fits in all three galaxies (see Table~\ref{tab:burkert_diagnostics}) and the residuals (inferred densities---B modeled densities) tend to remain within the sizes of observational uncertainties across the entire radial range (see Figures~\ref{fig:Burkertfits} and~\ref{fig:burkertfitsresiduals}).

\vspace{-6pt}

\begin{table}[H]
\caption{Median values of the Burkert density profile parameters and the corresponding virial mass for Draco, Sextans, and UMi dwarfs. Uncertainties correspond to $1\sigma$ credible intervals.}
\label{tab:burkert_parameters}
\setlength{\tabcolsep}{6pt}
\begin{tabularx}{\textwidth}{LCCC}
\toprule
\textbf{Parameter} & \textbf{Draco} & \textbf{Sextans} & \textbf{UMi} \\
\midrule
$\rho_{0}$ (g cm$^{-3}$)
& $(3.41^{+0.12}_{-0.11})\times10^{-23}$
& $(3.07^{+0.05}_{-0.05})\times10^{-24}$
& $(4.74^{+0.09}_{-0.09})\times10^{-24}$ \\
$r_{0}$ (kpc)
& $0.26^{+0.008}_{-0.008}$
& $1.17^{+0.03}_{-0.03}$
& $0.88^{+0.02}_{-0.02}$ \\
$M_{\mathrm{vir}}$ ($M_\odot$)
& $(4.04^{+0.23}_{-0.21})\times10^{8}$
& $(2.35^{+0.17}_{-0.14})\times10^{9}$
& $(1.67^{+0.09}_{-0.09})\times10^{9}$ \\
\bottomrule
\end{tabularx}
\end{table}

 In these objects, the (extrapolated) virial masses are in the range $ (4 \times 10^{8}- 2\times 10^{9})\ M_\odot $, while the stellar spheroid masses are in the range $(3-4) \times 10^5\ M_\odot$ (as estimated in \citep{masses}), i.e., they are >$2 \times 10^3$ times smaller than those of the dark halos, confirming that these systems are essentially spheres of dark matter.

\section {The Structure of the dSph Halos} \label{sec4}

 Let us point out that the DM halo density profiles that emerge in the most popular dark particle scenarios ($\Lambda CDM$, $\Lambda WDM$, Axion/Fuzzy DM, Self-Interacting DM) have two structural parameters that change object by object \citep{Salucci2019_DM_galaxies}. Similarly, the empirical DM density profiles have been found to have two  {{\bf SP}}s too that vary among objects \citep{Salucci2019_DM_galaxies}, providing us with a very valuable investigation of the scenarios above. Before discussing the results obtained in the Burkert framework, let us summarize the last 20 years of investigations of the dark halos in {\it  {disk systems}}. In these objects, the luminous and the dark matter components of the circular velocity are essentially distributed in an exponential thin disk of length scale $R_D$ and mass $M_D$ allied with a spherical halo with the density law of Equation \eqref {eq:BUR} \citep{Salucci2019_DM_galaxies}. Remarkably, and prospecting a physical origin, all the above four  {{\bf SP}} result %
  connected by ``fundamental'' relationships, i.e., each  {{\bf SP}} correlates with the other three \citep{S07}. In particular, the following relation is of importance for the present work (see \citep{Salucci2019_DM_galaxies} and references therein): 
\begin{equation}
 \log \Big(\frac{\rho_0 r_0}{M_\odot pc^2} \Big)= 2.15 \pm 0.2 
 \label{mu0}
\end{equation}
  {Equation}~\eqref{mu0} shows that, in galactic halos of very different masses, the central DM halo surface density $\propto \rho_0 
r_0$ is almost constant. In detail, Equation~\eqref{mu0} has been found to hold in disk systems ranging over five orders of magnitude in their virial  {mass} %
 $M_\mathrm{vir} $.
It is worth noting that Equation~\eqref{mu0} connects two quantities that do not exist in the NFW/cuspy scenario. We briefly remind the reader that in the B scenario, $\rho_0$ is defined as the value to which, moving from, e.g., $R_D$ towards the center, the halo density $\rho(R)$ {\it  {converges}}. In the NFW scenario, such a converging value does not exist: the DM density, in fact, diverges as $\propto R_D/R$, for $R \rightarrow 0$. This can be illustrated by noticing that, from $r=2\, R_D$ and $r= 0.2 \, R_D$, the NFW halo density increases by a factor of 10. Similarly, in the B scenario, one defines $r_0$ as the radius where the local DM density takes a value which is a fixed fraction of the {\it  {central}} value. In the NFW scenario, we do not have a central value for the DM density to determine this quantity.
Therefore, objects with statistically well-measured $r_0$ and $\rho_0$ strongly support the B scenario independently of whether, in a sample, these structural quantities are found to correlate among themselves. When they do so, they are indicating something more than the presence of a cored halo.

\begin{figure}[H]
 \includegraphics[width=0.65\textwidth]{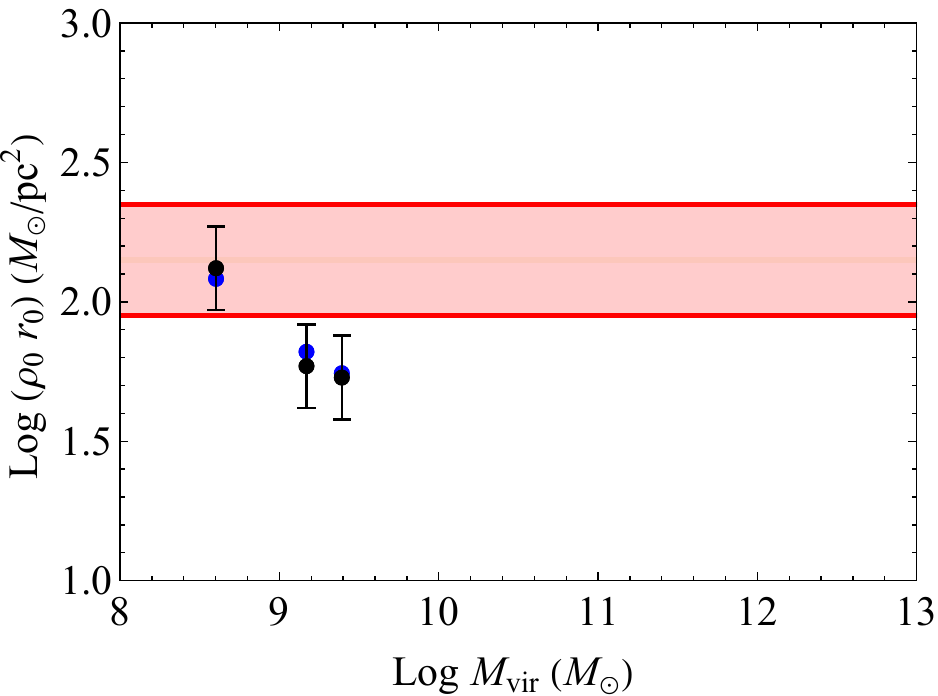}
 \caption{Central DM surface density $\rho_0 r_0$  {vs.} %
 galaxy mass $M_{\mathrm{vir}}$. Units: ($M_\odot/pc^2$,$M_\odot$). 
 The red band corresponds to the constant value ($\pm 1\sigma$) found in disk galaxies. The values of the dSphs are shown ( {black circles}). Blue circles refer to the B+NSC mass model of Section \ref{sec5}.
}
 \label{fig2new}
\end{figure}

This relationship has also been found to hold for a small sample of dSphs, although this has been investigated with a limited amount of kinematical data \citep{Salucci2012_dSph_scaling}. In this work, for each dSph, we have a large number of kinematical measurements that, in connection with a very accurate modeling procedure, have led to a robust determination of the Burkert  {{\bf SP}}s and, in turn, of the RHS of Equation~\eqref{mu0}. We find that the three dSphs follow the above disk-systems relationship quite well, as depicted in Figure~\ref{fig2new} and quantified in Table~\ref{tab:mu0_residuals} by normalized residuals. The normalized residual is defined as the difference between the measured value of the relation, in this case, $\log (\rho_0\; r_0 / (M_\odot\,\mathrm{pc}^{-2}))$, and the reference relation value $2.15 \pm 0.2$, divided by the total scatter. The quantity $\log (r_0 \rho_0)$ is found to be consistent with the relation within the total scatter of the disk-galaxy scaling law. On average, this ``constant'' quantity is found to be $\sim 1.4$ $\sigma $ (see Table \ref{tab:mu0_residuals}) lower than the value found in dIrr.
The small number of objects still limits us. It is relevant that, for the fully DM-dominated dSphs, the innermost regions of their DM halos can be represented as spheres of almost constant density: additionally, for the fully DM dominated dSphs, the innermost regions of their DM halos can be represented as spheres of almost constant density $\rho_0$ and of size $r_0$, with $\rho_0 =k /r_0$.

\begin{table}[H]
\caption{Normalized residuals ($N_{\sigma}$)
of the dSph galaxies relative to Equation~\eqref{mu0} for the two Burkert-profile models, expressed in units of the combined scatter.}
\label{tab:mu0_residuals}
\begin{tabularx}{\textwidth}{LCC}
\toprule
\textbf{Galaxy} &
\boldmath{$N_{\sigma}$} \textbf{(Burkert)} &
\boldmath{$N_{\sigma}$} \textbf{(Burkert + NSC Spike)} \\
\midrule
Draco & $-0.16$ & $+1.2$ \\
Sextans & $-2.12$ & $-0.9$ \\
UMi & $-1.80$ & $-0.8$ \\
\bottomrule
\end{tabularx}
\end{table}
\vspace{-6pt}

\begin{figure}[H]
 \includegraphics[width=0.65\textwidth]{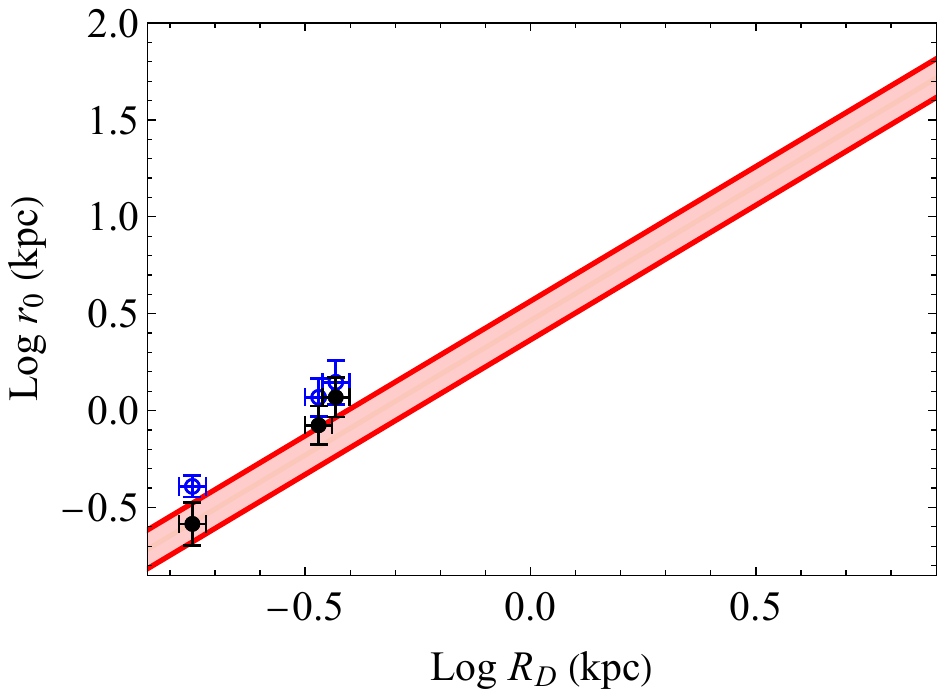}
 \caption{Relation between the dark matter core radius $r_0$ and the stellar disk scale length $R_D$. 
 The red band shows the scaling law for disk galaxies, while the blue/black points represent the three dSphs with(out) the central NSC of Section \ref{sec5}.
}
 \label{fig1new}
\end{figure}

Furthermore, in disk systems, including dIrrs and in M 87, a remarkable DM-LM scaling law emerges, relating the size $r_0$ of the DM halo constant density region with $ R_D$, the length scale of the stellar disk \citep{Donato2004_cores_scalelengths,S07, Salucci2019_DM_galaxies}:
\begin{equation}
 \log r_0 = 0.47 + 1.39 \log R_D.
\label{eq:length}
\end{equation}
with an r.m.s. of 0.05 dex. For the Plummer stellar spheroids of dSphs, $R_D^{sph}$, the length-scale equivalent to the disk $R_D$ is given by Equation~\eqref{eq:RDPLUMMER}. The three dSphs are found (see Figure~\ref{fig1new} and Table~\ref{tab:r0_residuals}) to very closely follow the same relationship held in disk systems, %
 despite their differences in morphology, dynamics, and mass. Quantitatively, all three systems, see Table \ref{tab:r0_residuals}, lie within approximately $1.5\sigma$ of %
  the dIrr relationship once the total scatter is taken into account, with Sextans and UMi showing particularly small residuals.
Remarkably, Equation~\eqref{eq:length} is valid in dSphs where, at any radius, the stellar component is negligible, and so is %
 the estimated number of exploded supernovae, making it impossible that baryonic feedback could have created such a relationship.\endnote{Also in disk systems in an assumed collisionless DM scenario, Equation~\eqref {eq:length} does not have a simple physical explanation. %
 }

\vspace{-6pt}

\begin{table}[H]
\caption{Normalized residuals ($N_{\sigma,r_0}$) of the dSphs relative to Equation~\eqref{eq:length} for the two Burkert-profile models, expressed in units of the combined scatter. $N_{\sigma,r_0}$ measures the difference between the observed $\log r_0$ and the value predicted by Equation~\eqref{eq:length}, normalized by the total scatter.}
\label{tab:r0_residuals}
\begin{tabularx}{\textwidth}{LCc}
\toprule
\textbf{Galaxy} &
\boldmath{$N_{\sigma,r_0}$} \textbf{(Burkert)} &
\boldmath{$N_{\sigma,r_0}$} \textbf{(Burkert + NSC Spike)} \\
\midrule
Draco & $-1.4$ & $+0.4$ \\
Sextans & $-0.3$ & $+0.1$ \\
UMi & $-0.6$ & $- 0.0$ \\
\bottomrule
\end{tabularx}
\end{table}

\section{The Innermost Regions of dSphs: Intermediate-Mass Black Holes, Nuclear Stellar Cluster?}\label{sec5}

 The three dSphs show a cored DM density region of size $r_0 =(260-1200)$ pc.
 In Figure~\ref{fig:Burkertfits}, by comparing the inferred DM halo densities $\rho(r_k)$ and the best-fit B densities $\rho_B(r)$, one realizes that, moving from $R_o$ towards the galaxy center, both densities converge to a constant value $\rho_0$ at $r = R_C \sim (3/4-1/3)\ r_0$, with $R_C=(200-400)$ pc
 (see Table~\ref{tab:inner_radii}).
 Continuing in moving inwards to the galaxy center, as expected, $\rho_B(r)$ remains constant down to $r = R_i$, while $\rho(r_k)$ starts to increase and reaches, at $r _k= R_i$, a value which is a couple of times bigger than $\rho_0$, with an estimated significance of $\sim$1.5 $\sigma$ (see Table~\ref{tab:inner_radii}). 

 This feature could indicate the presence of a central massive component; in the 3 dSphs, inside the radii $R_C$, one finds excesses in mass of the order of $M_C= [5,1,2]\times (1\pm 0.5) \times 10^6M_\odot $ 
(see Table~\ref{tab:inner_radii}) worth comparing with the corresponding halo virial masses $M_\mathrm{vir}=$$ [0.4,2.3,1.7] \times (1\pm 0.1) \times 10^9\, M_\odot $, see Table \ref{tab:burkert_parameters}. These features present in the innermost halo regions, which, tentatively, we call ``spikes'', are therefore of modest size %
 and not fully statistically confirmed. However, they are relevant in that, if eventually proven, they do not have an explanation in terms of the various DM halo density profiles considered so far. On the other hand, the values of $\rho_k(r_k)$ in the region between $R_i$ and $R_C$ provide us with upper limits on the dark halo densities and, in turn, on the mass of a (hypothetical) very central massive component. 

\begin{figure}[H]
 \includegraphics[width=0.45\textwidth]{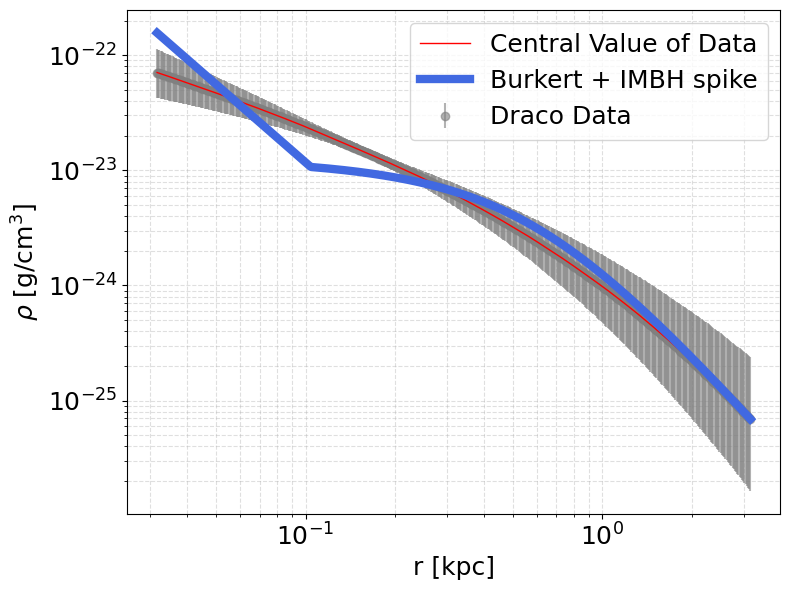}
 \includegraphics[width=0.45\textwidth]{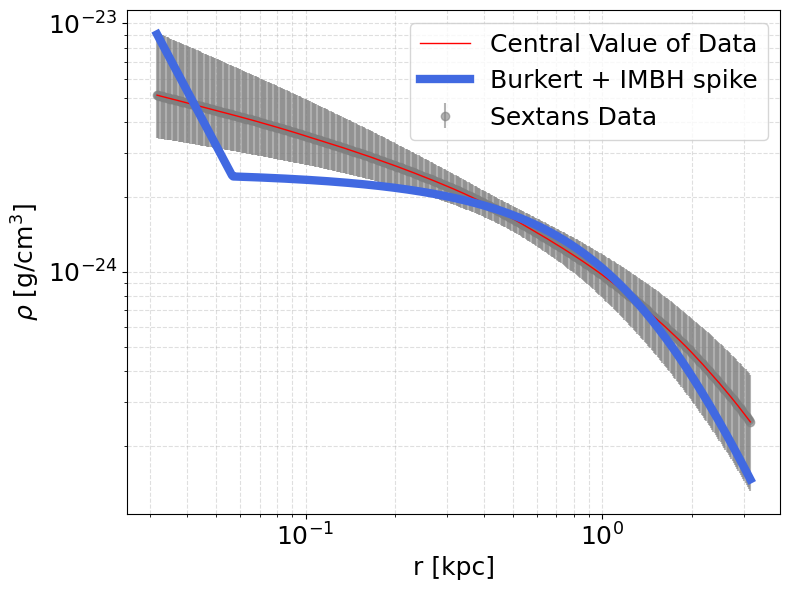}
 \includegraphics[width=0.45\textwidth]{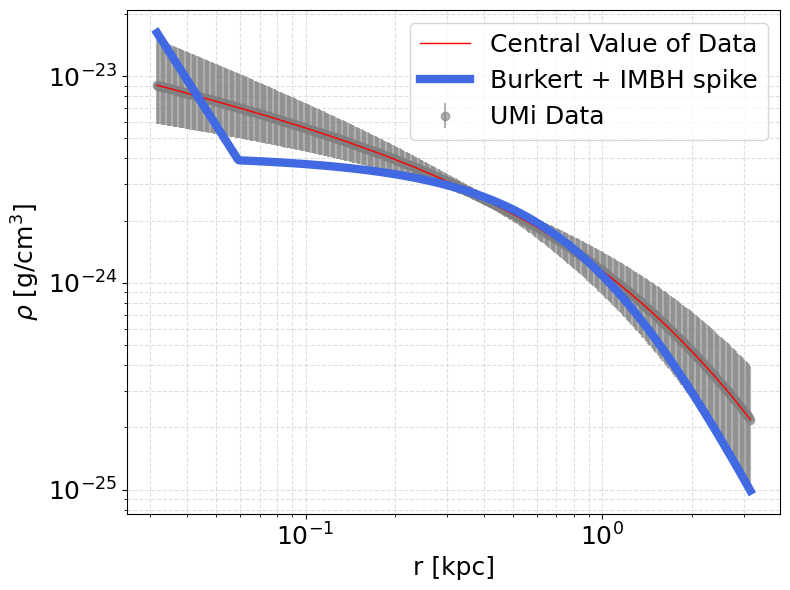}
 \caption{Dark matter density profiles of Draco, Sextans, and UMi compared with their best-fitting Burkert + IMBH spike fit ($g=-2.25$) model. Gray points denote the inferred densities with uncertainties, and the blue curves show the corresponding fits. The red curves connect the central values of the measurements.}
 \label{fig:IMBHfits}
\end{figure}

Are these ``spikes'' related to the ones that get dynamically created in a DM halo by the presence of an intermediate-mass %
 black hole (IMBH) of mass $M_\bullet\sim 10^{5-6} M_\odot$, located at the center of a galaxy like the present ones \citep{Gondolo1999, Sagittarius,IMBH2,IRA}? To analyze this possibility, we take the working assumption that, in the three dSphs, the innermost regions of size $R_C\sim 200$ pc are gravitationally dominated by a central IMBH, which causes, there, the failure of the B profile in representing the inferred halo densities $\rho(r_k)$ (see Figure~\ref{fig:Burkertfits}). Then, the presence of the spike in the DM halo density is considered as the signature of an IMBH at its center. It is well-known that, given a central IMBH of mass $M_\bullet$, for $ r\leq R_{sp}$, with the latter radius defined by
 \begin{equation}
 4\pi \int _0^{5 \,R_{sp}} r^2 \rho_B(r_; r_0,\rho_0)\, dr=2 M_\bullet.
 \label{bhmass}
 \end{equation}
 a spike in the DM distribution emerges, and the halo density profile gets modified to~\citep{Gondolo1999, Kame}: 
\begin{equation} 
\rho_\text{IMBH}(r;r_0,\rho_0,R_{SP})=\rho_B(r; r_0;\rho_0) \ \ \ \ r\geq R_{sp}
\label{rhosp0}
\end{equation}\begin{equation}
\rho_\text{IMBH}(r;r_0,\rho_,R_{SP}) =\rho_{sp} {\Big (} \frac{r}{R_{sp}} {\Big)}^{g} \ \ \ \ r\leq R_{sp} 
\label{rhosp}
\end{equation}
with $g=-2.25$, $\rho_{sp} =\rho_B(R_{sp}, r_0; \rho_0)$. From Equations~\eqref{bhmass} and \eqref{rhosp} we have $ R_{sp}=\Big(\frac {0.75 \, 0.2^{0.75} M_\bullet}{2 \pi \rho_{sp}}\Big)^{1/3}$. We then proceed by fitting $\rho(r_k)$ with the B+IMBH model (built by combining Equations~\eqref{bhmass}--\eqref{rhosp}) and we obtain the best-fit values: $ r_0, R_{sp} $, $\rho_{sp}$ and, in turn, the IMBH mass $M_\bullet$. The results are shown in Figure~\ref{fig:IMBHfits} and in Table~\ref{tab:parameters_imbh}. We realize that, in all galaxies, the above excess $\rho(r_k)-\rho_B(r_k)$ emerging inside $R_C$ cannot be due to the existence of a central IMBH. In the three objects, in the region between $R_i$ and $R_c$, the value of the power law slope of the inferred density %
$\rho(r_k)$ is: $\Gamma \equiv d\log\, \rho(r_k)/d\log \, r_k$ is certainly much shallower than the value $g=-2.25$ of the B+IMBH scenario. The values of the fractions $M_\bullet/M_{sph}$ are %
 much bigger than the expected ones: $\sim$$10^{-2}$--$ 10^{-3}$ (see Table~\ref{tab:parameters_imbh}).
 In the three dSphs, therefore, we can exclude that, for $r\geq R_i$, the tracer stars are subjected to the gravitational field of a central IMBH; our mass model sets an upper limit for the mass of such a component of $M_\bullet \leq [24,1,2.5] \times 10^5\, M_\odot $, see Figure~\ref{fig:IMBHfits} and Table \ref{tab:parameters_imbh}.

\begin{figure}[H]
 \includegraphics[width=0.45\textwidth]{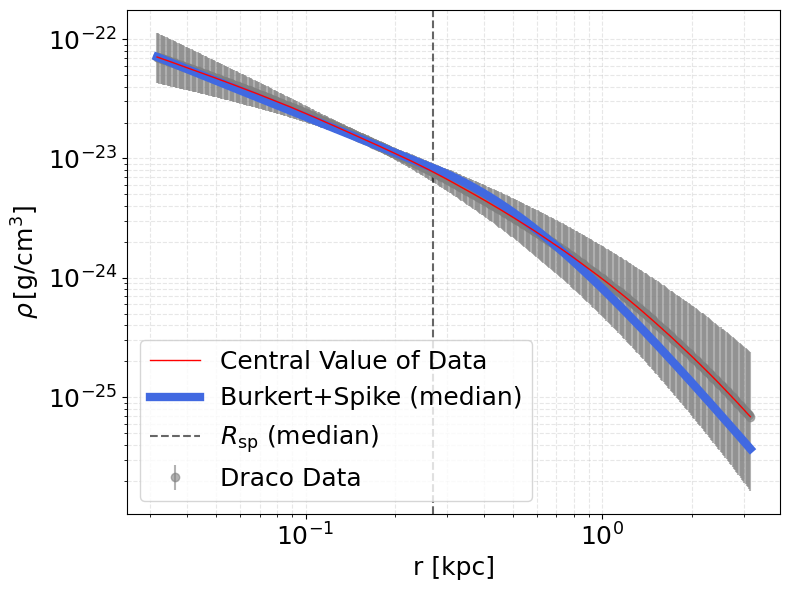}
 \includegraphics[width=0.45\textwidth]{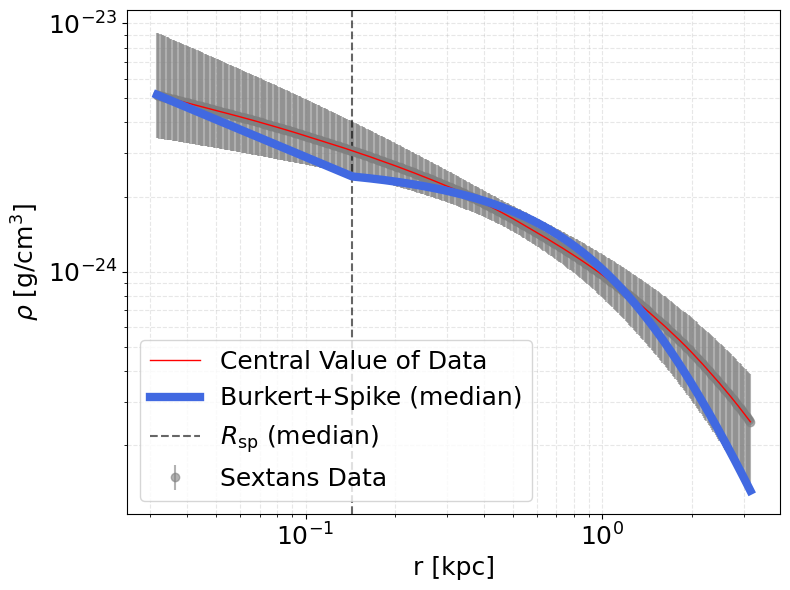}
 \includegraphics[width=0.45\textwidth]{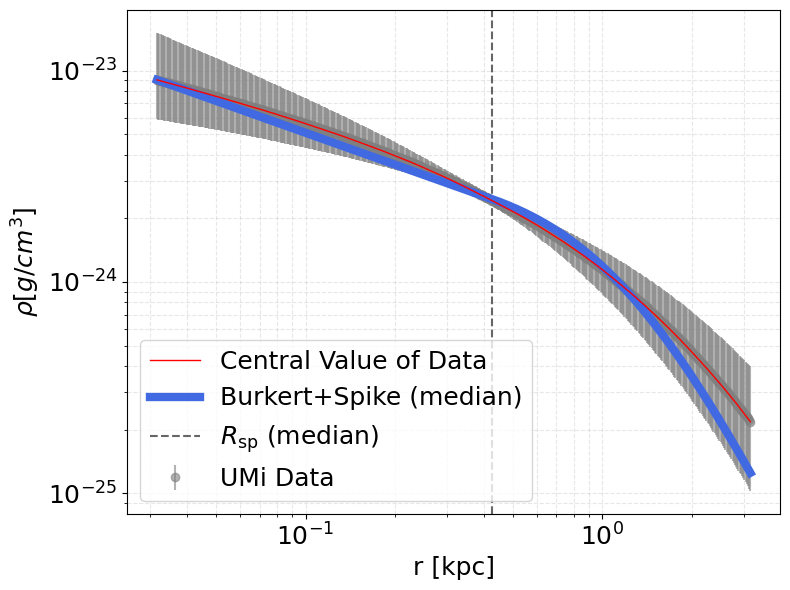}
 \caption{
 Dark matter density profiles of Draco, Sextans, and UMi compared with their best-fitting Burkert + IMBH spike fit ($g=-1$ for Draco and $g=-0.5$ for Sextans and UMi). Gray points denote the inferred densities with uncertainties, and the blue curves show the corresponding fits. The red curves connect the central values of the measurements.}
 \label{fig:ghalffits}
\end{figure}

 A second possibility to explain the above ``spikes'' is the presence, inside the innermost $\sim$100  pc of the dSphs, of a stellar component, called a nuclear star cluster \citep{Poulain,Sancheznsc,Sagittarius}, with mass in the range $ (10^5- 10^7)M_\odot $ and a value of its power law density slope $g$ in the range $-1 \leq g \leq-0.5 $, which is  %
  more in line with the observed values of $\Gamma $. Again, we assume that the presence of a ``spike'' = the presence of an NSC, and we investigate the dSph's inferred density in this frame. The results of the best fit of the B+NSC model are shown in Figure~\ref{fig:ghalffits} and in Table~\ref{tab:posterior_params_and_Mbh_rot_sym_fmt}. We find the presence inside $R_C$ (or $R_{SP}$) regions mapped by the $\rho(r_k)$s of Matrix A of an NSC of mass $ M_{NSC}=(5 \times 10^5-10^7)\, M_\odot$  {(}%
Table~\ref{tab:posterior_params_and_Mbh_rot_sym_fmt}).

We notice that, by adding in the density model the presence of a central NSC, this somewhat decreases the values of the residuals, i.e., (predictions - measurements) %
 (see Tables~\ref{tab:mu0_residuals} and \ref{tab:r0_residuals}), but does not affect the relationships among the galaxies' structural parameters (see Figures~\ref{fig2new} and \ref{fig1new}) found with the simplest B density model.

 The issue with this scenario is that the mass of the NSCs results one order of magnitude bigger than that of the stellar component of the whole dSph as evaluated from the galaxy photometry in \citep{masses} and reported in Table~\ref{tab:inner_radii}.\endnote{One could hypothesize that, inside $R_C$, the ``stellar'' mass to light ratio is much higher than in the outer parts of the stellar spheroids due to the presence there of a certain number of stellar BHs of mass > $100\ M_\odot$ formed during the complex dynamical stages that this region has undergone.}

\begin{table}[H]
\caption{Here, the following is tabulated: length scale of a Plummer sphere corresponding to the Freeman disk length scale $R_D$, the distance $D$ of the dSph, the innermost radius $R_C$ at which the B model fits the data, 
the increase in density
between $R_C$ and the innermost radius $R_i$: $\Delta \rho = Log[\rho(R_C)] - Log[\rho(R_i)]$, the stellar mass of the galaxy $M_\star$ (with an estimated uncertainty of 30 \%), the mass inside $R_C$, $M_C$ (with an estimated uncertainty of 50\%).}
\label{tab:inner_radii}
\begin{tabularx}{\textwidth}{L lc c c c c}
\toprule
\textbf{Name} & \boldmath{ {$R_D^ {sph}$} %
} \textbf{[kpc]}& \textbf{\boldmath{$D$} [kpc]}& \textbf{\boldmath{$R_C$} [kpc]} & \textbf{\boldmath{$\Delta \rho$} [dex]} & \boldmath{$M_\star$ [$10^5M_\odot$]}& \boldmath{$M_C$ [$10^6M_\odot$]}\\
\midrule
Draco & $0.277$ & 75.8 & $0.12$ & 0.5 &2.9 &5.0\\
Sextans & $0.577$ & 92.5& $0.19$ & 0.2 &4.4 &1.1\\
UMi & $0.529$ & 76.4& $0.22$ & 0.5 &2.9&2.0\\
\bottomrule
\end{tabularx}
\end{table}
\vspace{-6pt}

\begin{table}[H]
\caption{Median values of the Burkert density profile parameters, including an IMBH-induced density spike ($g=-2.25$) for Draco, Sextans, and UMi dwarf spheroidal galaxies. Quoted uncertainties are symmetrized $1\sigma$ errors.}
\label{tab:parameters_imbh}
\setlength{\tabcolsep}{8pt}
\begin{tabular}{lccc}
\toprule
\textbf{Parameter} & \textbf{Draco} & \textbf{Sextans} & \textbf{UMi} \\
\midrule
$\rho_{0}$ (g\,cm$^{-3}$)
& $(3.00 \pm 0.10)\times10^{-23}$
& $(3.01 \pm 0.05)\times10^{-24}$
& $(4.57 \pm 0.09)\times10^{-24}$ \\
$r_{0}$ (kpc)
& $0.292 \pm 0.009$
& $1.196 \pm 0.033$
& $0.923 \pm 0.025$ \\
$R_{\mathrm{sp}}$ (kpc)
& $0.063 \pm 0.002$
& $0.042 \pm 0.003$
& $0.052 \pm 0.003$ \\
$M_{\mathrm{vir}}$ ($M_\odot$)
& $(4.82 \pm 2.63)\times10^{8}$
& $(4.30 \pm 1.52)\times10^{9}$
& $(2.12 \pm 1.10)\times10^{9}$ \\
$M_{\bullet}$ ($M_\odot$)
& $(2.43 \pm 0.22)\times10^{6}$
& $(9.19 \pm 1.49)\times10^{4}$
& $(2.54 \pm 0.34)\times10^{5}$ \\
\bottomrule
\end{tabular}
\end{table}
\vspace{-6pt}

\begin{table}[H]
\caption{Median values of the Burkert density profile parameters, including an NSC density spike for the Draco, Sextans, and UMi dwarf spheroidal galaxies. Quoted uncertainties are symmetrized $1\sigma$~errors.}
\label{tab:posterior_params_and_Mbh_rot_sym_fmt}
\setlength{\tabcolsep}{8pt}
\begin{tabular}{lccc}
\toprule
\textbf{Parameter} & \textbf{Draco} & \textbf{Sextans} & \textbf{UMi} \\
\midrule
$g$
& $-1.0$
& $-0.5$
& $-0.5$ \\
$\rho_{0}$ (g\,cm$^{-3}$)
& $(1.99 \pm 0.07)\times10^{-23}$
& $(2.66 \pm 0.32)\times10^{-24}$
& $(3.80 \pm 0.04)\times10^{-24}$ \\
$r_{0}$ (kpc)
& $0.406 \pm 0.013$
& $1.40 \pm 0.05$
& $1.17 \pm 0.02$ \\
$R_{\mathrm{sp}}$ (kpc)
& $0.270 \pm 0.022$
& $0.145 \pm 0.080$
& $0.426 \pm 0.005$ \\
$M_{\mathrm{NSC}}$ ($M_\odot$)
& $(7.51 \pm 1.90)\times10^{6}$
& $(3.89 \pm 0.68)\times10^{5}$
& $(1.01 \pm 0.18)\times10^{7}$ \\
\bottomrule
\end{tabular}
\end{table}

\vspace{-6pt}

\vspace{-6pt}

\section{Discussion}\label{sec6}

Let us notice that the relationships in Equations~\eqref{mu0} and~\eqref{eq:length} are not related to the (possible) presence of deep inner ``spikes''.
Moreover, in the case in which the latter are caused by some random or systematic errors in the determination of the DM density in the regions $r<R_C$, Equation~\eqref{eq:jintegral} would still be valid in computing the actual J factor of the DM halos. 

It is then interesting to investigate the effect of the B+NSC mass model on the estimate of the J-factor that quantifies the astrophysical contribution to the expected signal from the dark matter (DM) annihilation in a galaxy. Let us assume, as in \citep{yang2025darkmattercontentmilky}, an angular integration region of 0.5° for each dSph at distance D, so we have
\begin{equation}
R_{\max}
=
 D \, \frac{0.5 \ \pi}{180}
 \end{equation}
\begin{equation}
J
=
\frac{4\pi}{D^{2}}
\int_{0}^{R_{\max}}
\rho_{B+NSC}^{2}(r)\, r^{2}\, dr
\label{eq:jintegral}
\end{equation}

 We realize (see Table \ref{tab:j_factors}) that this factor is smaller by 0.2 dex than the value obtained by assuming the gNFW profile because, in this model, the innermost region of 50--100 pc is filled with luminous matter rather than being a dark ``spike''.

\begin{figure}[H]
\includegraphics[width=0.65\textwidth]{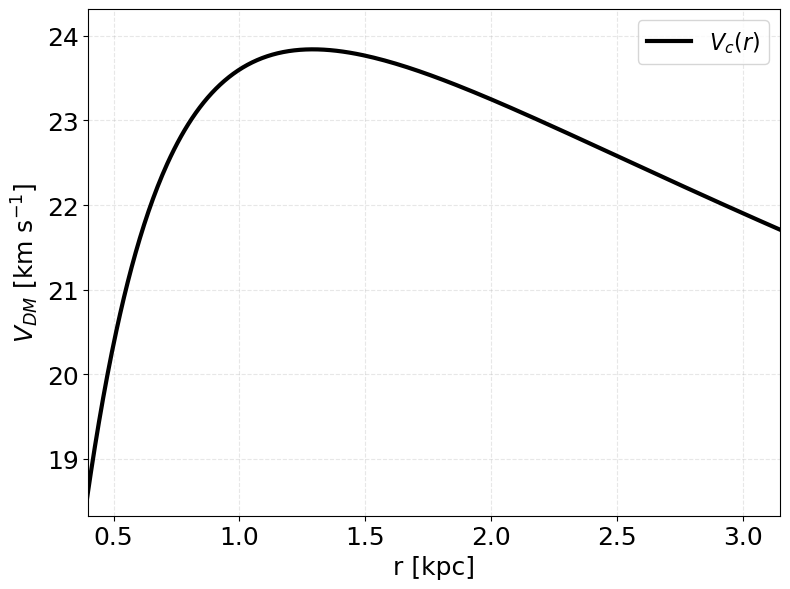} 
\caption{ {The} %
 contribution of the DM halo to the circular velocity $V$ in Draco. Data  reach the declining region of $V(r)$, which is %
 unavoidable in the dark particle + Newtonian Scenario. This shows a counterexample to the claim that the  RCs remain %
  asymptotically flat, as predicted by a number of alternatives to the dark particle. }
\label{fig4new}
\end{figure}

Let us point out that, in dSphs, we can directly investigate the DM density at larger distances, in terms of its halo length scale $r_0$, than those investigated in normal spirals, i.e., out to $r=R_o = (2\ r_0 - 8\ r_0)$. The radii investigated with available kinematics are well beyond the constant density region, which highly supports its detection. Moreover, in Draco, the kinematics reach the region where the DM halo contribution to the circular velocity is declining with radius (see Figure~\ref{fig4new} {)}%
. We can rule out, with this, the MOND prediction of a constant circular velocity out to a galaxy's outermost kinematical measurement.

\begin{table}[H]
\caption{Integrated $J$-factors within an angular radius of $0.5^\circ$. 
The $J$-factors in the second column are adapted from \citep{yang2025darkmattercontentmilky}, assuming a gNFW density profile, and
the third column reports $J$-factors obtained using the B+NSC model.}
\label{tab:j_factors}
\begin{tabularx}{\textwidth}{l C C }
\toprule
\textbf{Object} 
& \boldmath{$\log_{10} J_{\rm gNFW}(0.5^\circ)\;[\mathrm{GeV}^2\,\mathrm{cm}^{-5}]$}
& \boldmath{$\log_{10} J_{\rm B+NSC}(0.5^\circ)\;[\mathrm{GeV}^2\,\mathrm{cm}^{-5}]$}
 \\
\midrule
Draco & $18.74^{+0.12}_{-0.09}$ & $18.62 \pm 0.16$ \\
Sextans & $17.98^{+0.12}_{-0.12}$ & $17.83 \pm 0.16$\\
UMi & $18.27^{+0.07}_{-0.07}$ & $18.07 \pm 0.08$ \\
\bottomrule
\end{tabularx}
\end{table}

\vspace{-6pt}

\section{Conclusions}\label{sec7}

dSphs are among the oldest galaxies in the local universe, and are made by $10^{(5-6)}$ stars situated at the center of the smallest and densest dark halos in the present universe with virial masses of $(10^8-10^9)\ M_\odot $. They are excellent laboratories for investigating the nature of the dark particle and the formation of the first objects in the universe. 
The density measurements of \citet{yang2025darkmattercontentmilky}\endnote{See Footnote 1.} endow us with precious information on the properties of the DM in these objects and a meaningful clarification about the core/cusp character of their dark halos. Based on a large amount of stellar kinematical and photometric data of three nearby dSphs and by performing on each object a dynamical analysis through the framework of the ``neutral'' gNFW DM halo profile,  {ref.} %
 \citep{yang2025darkmattercontentmilky} derived dark halo densities (and their uncertainties) at 500 radii, situated between 32 pc and 3.2 kpc, Then, they determined the innermost radial slopes and found that these galaxies are essentially dark spheres. We have exploited these 
measurements to quantify the extent to which these halos are concordant/discordant with the full/recent predictions/observations of the NFW $\Lambda$CDM/Burkert halo profiles and whether they hint at the presence of an additional spherical component more compact than a dark halo.

In each object, we found that the inferred densities of the ``dark spheres'' are inconsistent with the NFW halo density profiles and with the related \textit{c} vs. $M_\mathrm{vir}$ cosmological relationship. The corresponding values of the SPs are so unrealistic that we do not see how a $\Lambda$CDM scenario could be restored in the three dSphs. Moreover, in these objects, the lack of a relevant stellar component supports the investigation of the DM density profile with an accurate new method that will independently validate the strong disagreement between the inferred and the NFW-predicted DM density profiles 

As far as the cored Burkert profile, we have  %
 that in this framework, in disk systems, a family of relationships among the various dark and luminous structural parameters emerges. We have tested the inferred dSph densities against such a profile and with respect to the above family of relationships. The Burkert profile fits the dSph densities well and, remarkably, the resulting structural parameters $R_D^{sph}, r_0,\rho_0$ correlate among themselves via the same relationships in spirals, including dIrrs. This is likely to signal the occurrence of some ``physical process'' between the dark particles and the (luminous) standard model particles. Specifically, the involvement of the dSphs in the relationships of Equation~\eqref{eq:length} (see Figure~\ref{fig1new}) is difficult to understand without resorting to some ``collisional'' physics for the dark particle.

In the very innermost regions of these objects, from $r =(0.1-0.2) \ Kpc $ down to $R_i= 0.04\ kpc$, the flat B halo density model profile is in some tension with the moderate increase shown by the inferred densities $\rho(r_k)$. However, we have found that such a feature cannot be due to a central intermediate-mass black hole nor to a central nuclear stellar cluster; to be understood, it needs further investigation beyond the determination of the upper limits of the mass of these two components.

\vspace{6pt}
\authorcontributions{ {~} %
Conceptualization, R.B. and P.S.; methodology, R.B. and P.S.; software, R.B. All authors have read and agreed to the published version of the manuscript.
}

\funding{ {~} %
This research was funded by INFN, Iniziativa Specifica QGSKY, CSN4---Fisica Teorica.
}

\dataavailability{ {~} %
The DM density profiles analyzed in this study are publicly available on Zenodo at \url{https://zenodo.org/records/16919342}. The code used for the fits is available at \url{https://github.com/rojinbayati/dm-mcmc-fits}.
 }

\conflictsofinterest{ {~} %
The authors declare no conflict of interest.
} 
 
\appendix
\appendixstart

 \section{~}
\begin{figure}[H]
 \includegraphics[width=0.47\textwidth]{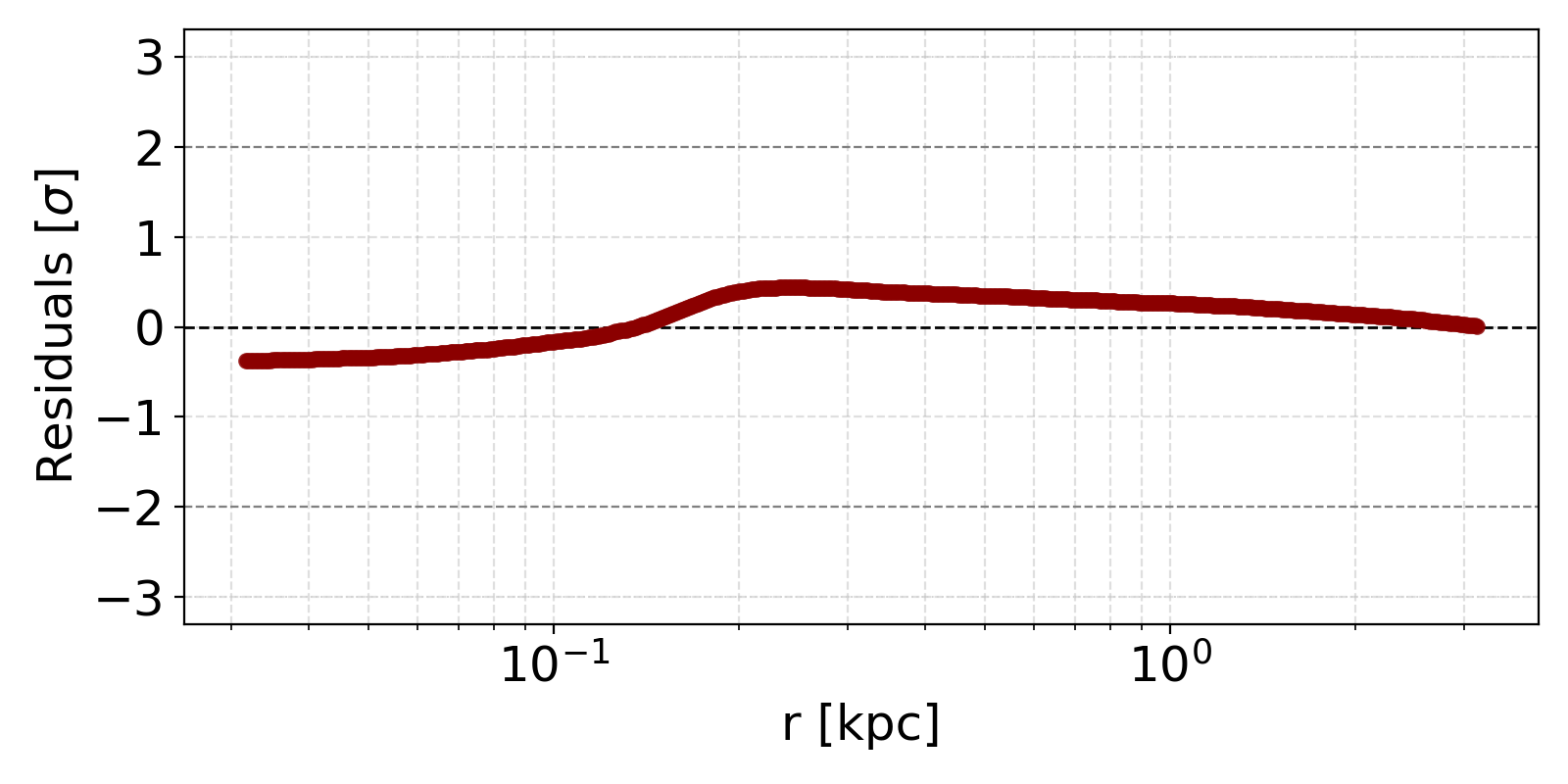}
 \includegraphics[width=0.47\textwidth]{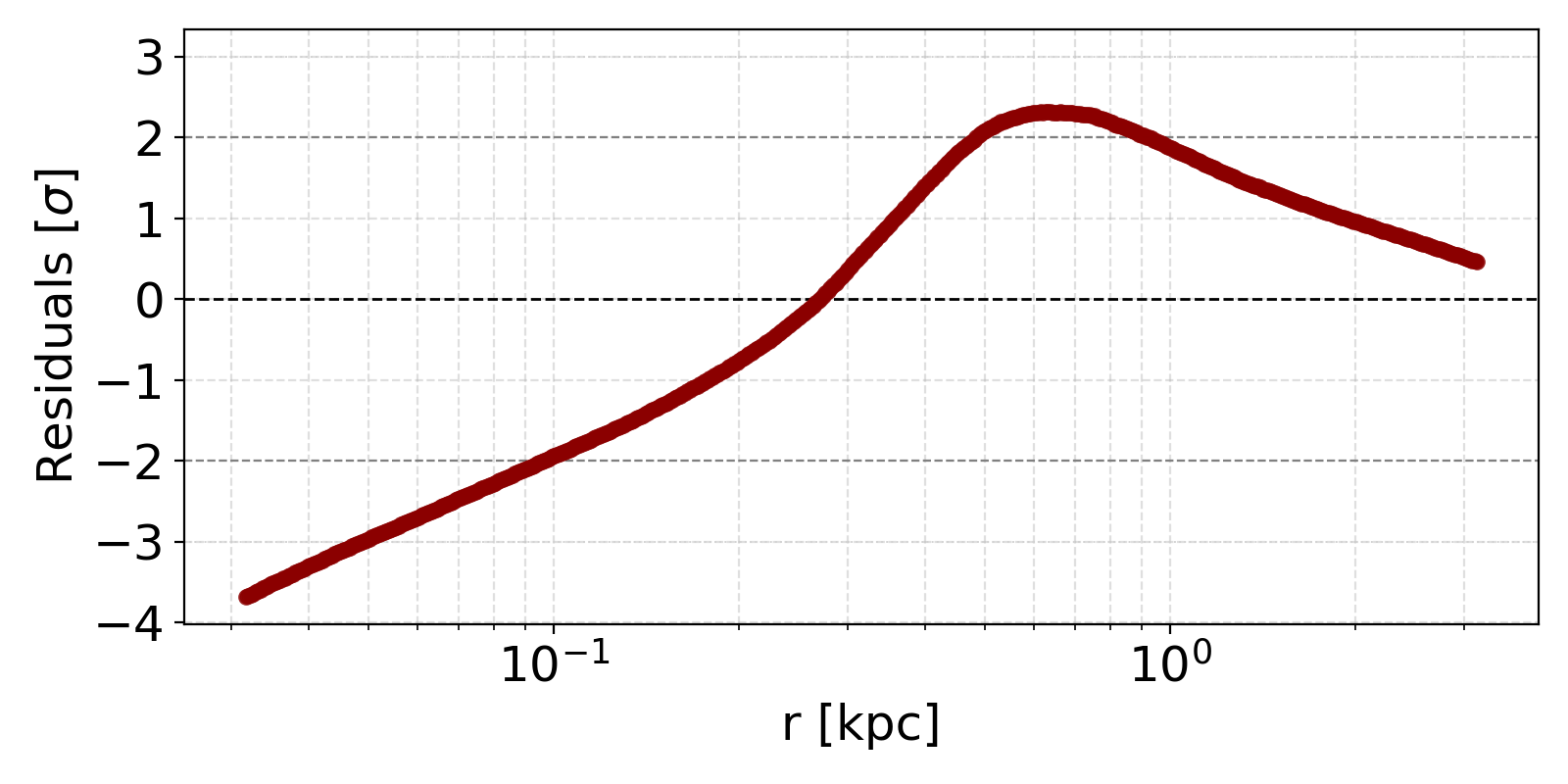}
 \includegraphics[width=0.47\textwidth]{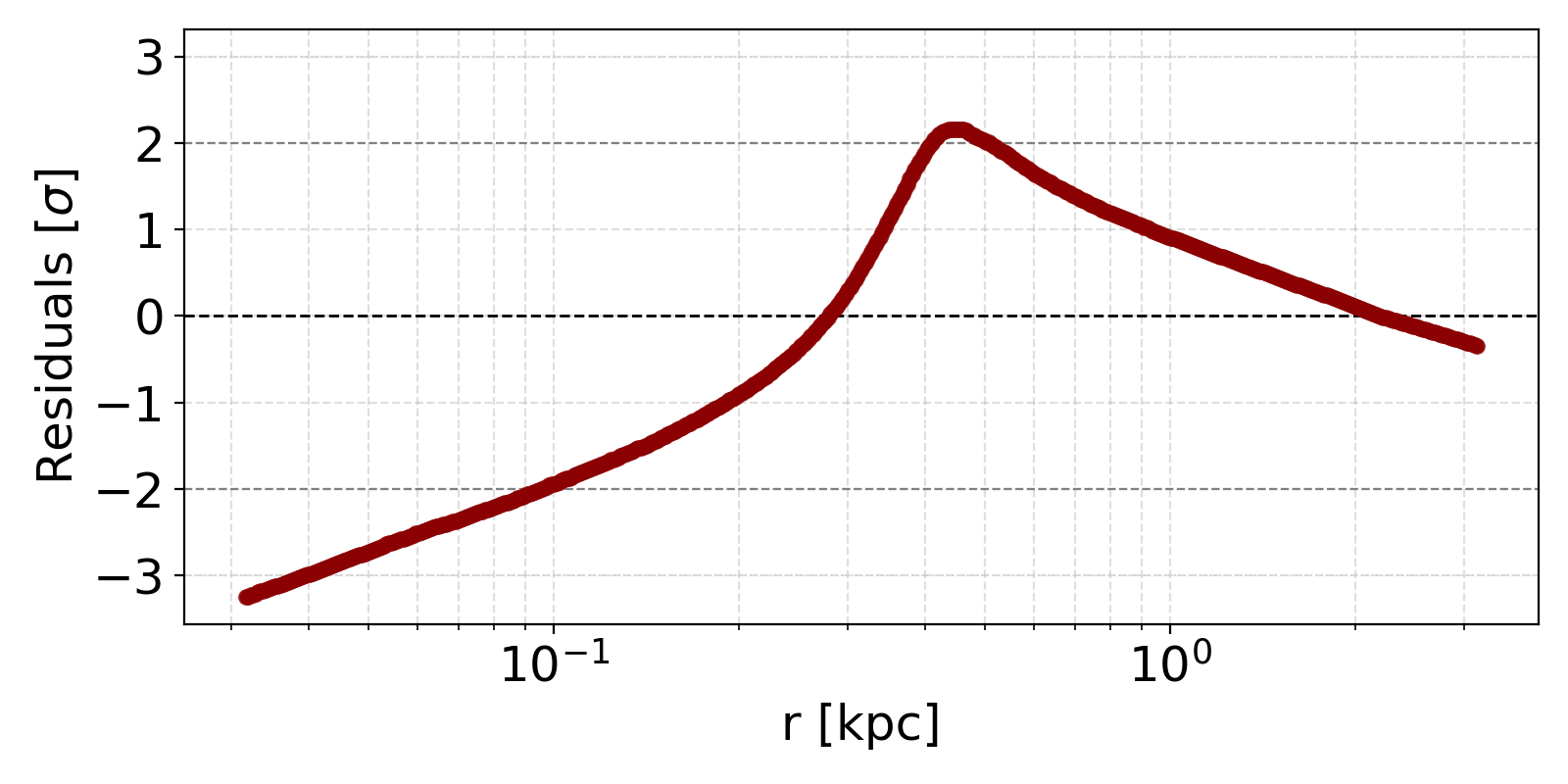}
 \caption{Residuals for NFW fit of Draco (top-left), Sextan (top-right) and UMi (bottom).}
 \label{fig:NFWfitsresiduals}
\end{figure}

 We present diagnostic measures for the fits obtained by assuming the NFW and the B density profiles. This is also  %
 evident in Figure~\ref{fig:NFWfits}, where the residuals are plotted as a function of radius in Figure~\ref{fig:NFWfitsresiduals} and the values of the reduced $\chi^2$ in Table~\ref{tab:nfw_diagnostics} show that the NFW profile provides a poor fit to the inferred halo densities of Sextans and UMi, while it effectively reproduces those of Draco (at the expense of taking nonphysical values for its halo structural parameters).
\begin{table}[H]
\small
\caption{Diagnostics of the NFW fits for Draco, Sextans, and UMi dwarfs.}
\label{tab:nfw_diagnostics}
\begin{tabularx}{\textwidth}{l CCC}
\toprule
\textbf{Diagnostic} & \textbf{Draco} & \textbf{Sextans} & \textbf{UMi} \\
\midrule
Mean acceptance fraction 
& 0.68 
& 0.65 
& 0.53 \\
Reduced $\chi^2$ 
& 0.08 
& 3.64 
& 2.76 \\
\bottomrule
\end{tabularx}
\end{table}
\vspace{-8pt}

\begin{figure}[H]
 \includegraphics[width=0.48\textwidth]{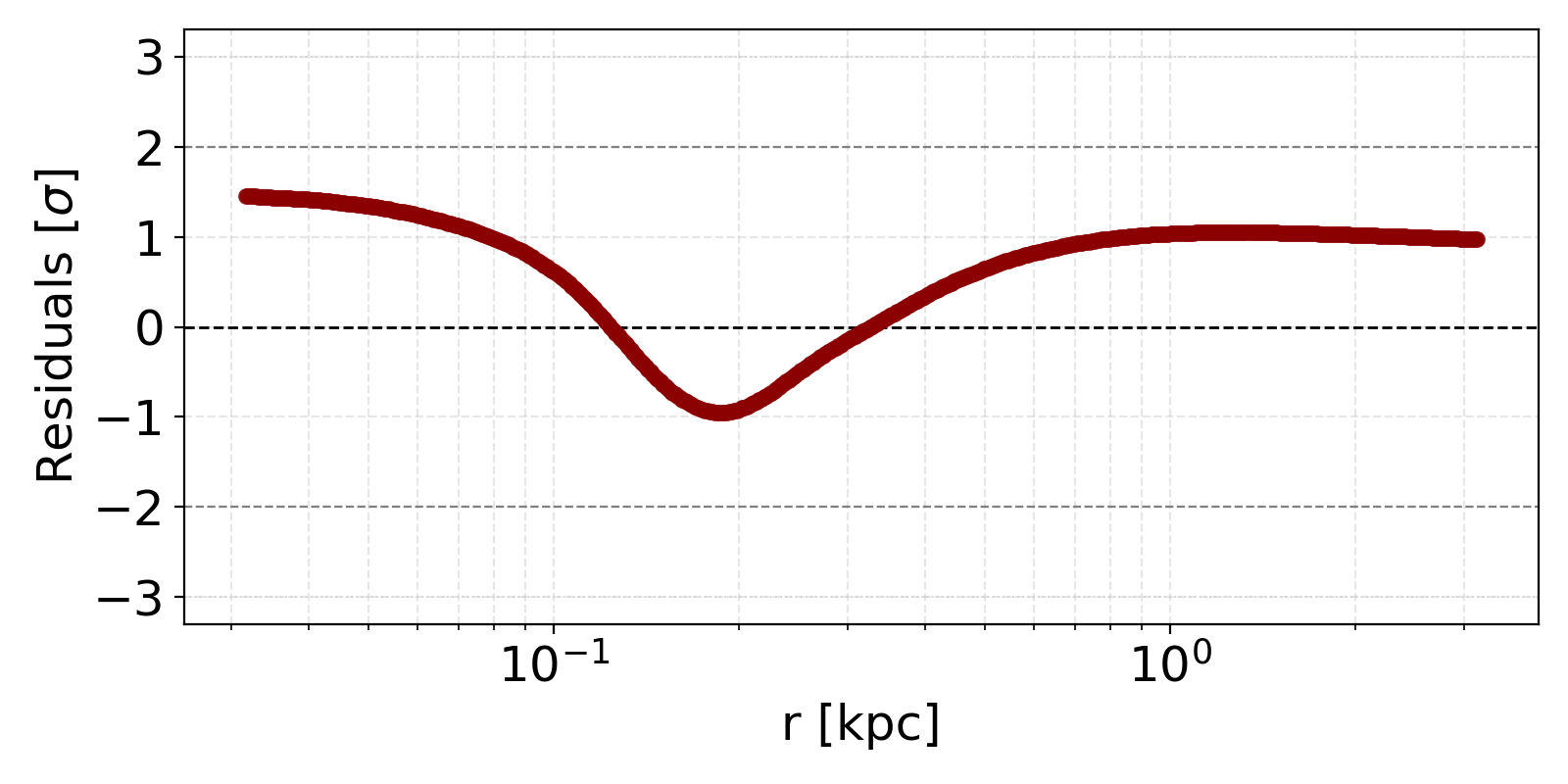}
 \includegraphics[width=0.48\textwidth]{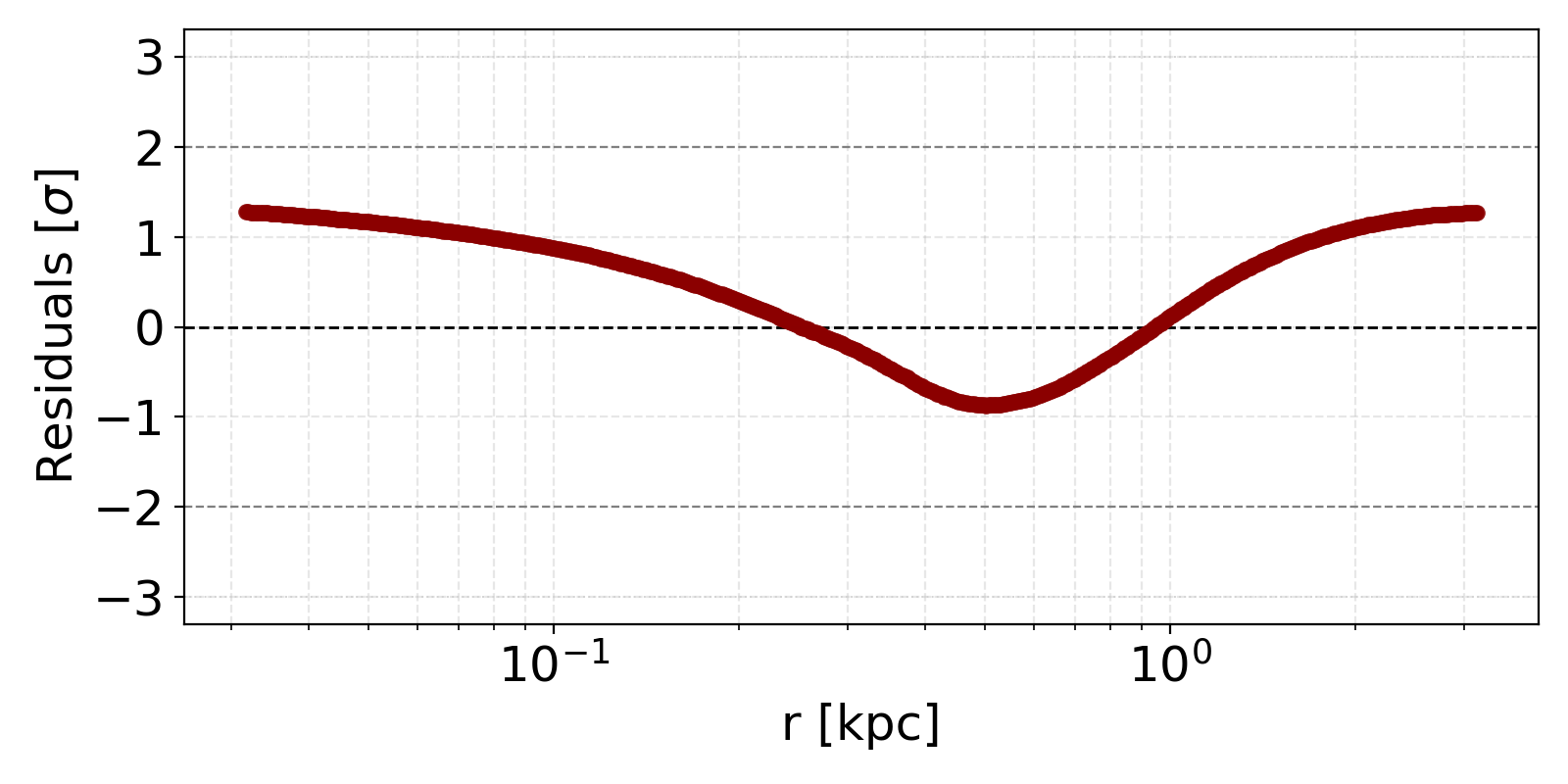}
 \includegraphics[width=0.48\textwidth]{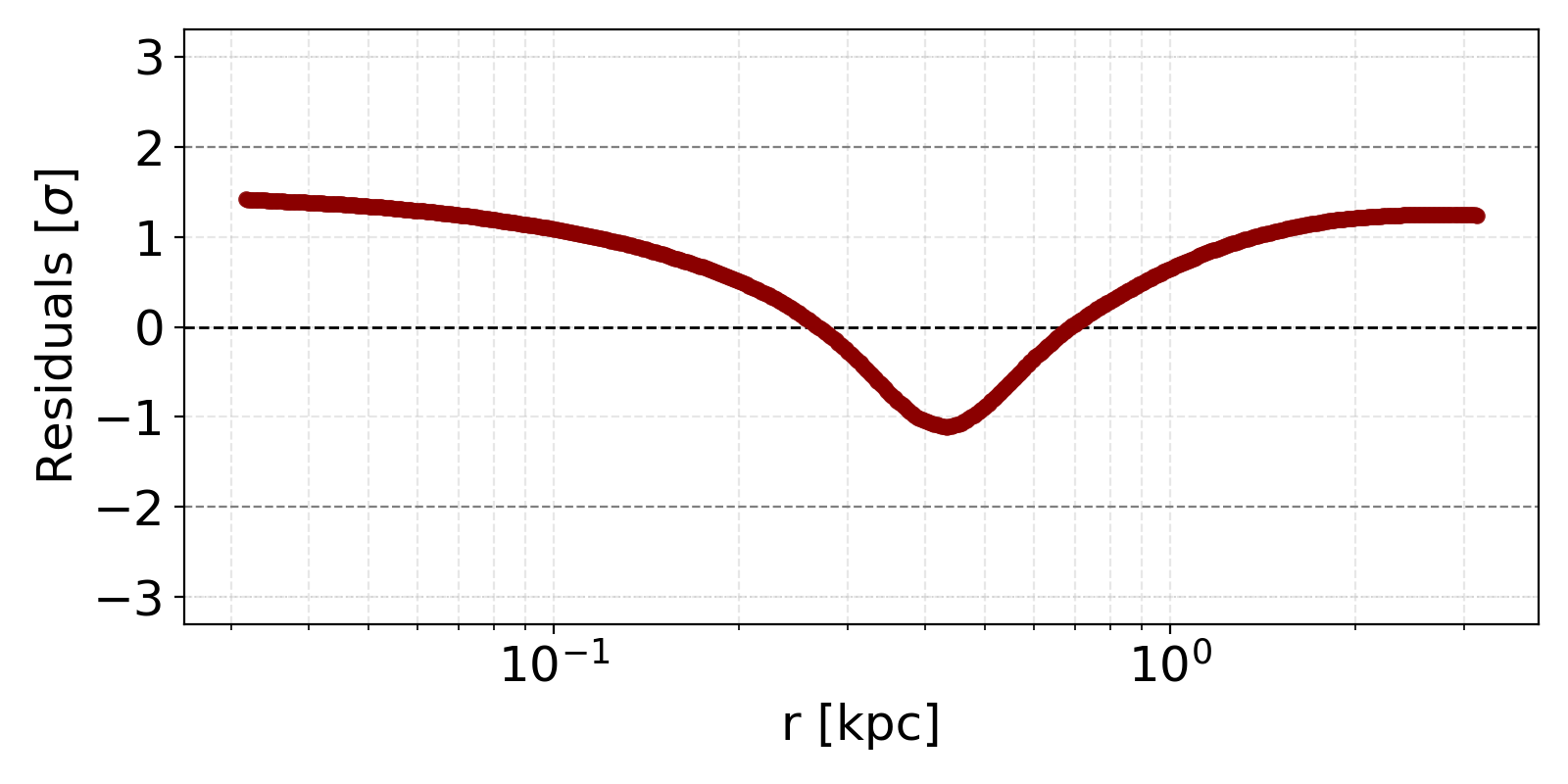}
 \caption{Residuals of the Burkert fit of Draco (top-left), Sextan (top-right) and UMi (bottom).}
 \label{fig:burkertfitsresiduals}
\end{figure}

By applying the same statistical analysis by assuming a B halo model, we realize that the latter obtains a good-fitting performance in all three dSphs (see Table~\ref{tab:burkert_diagnostics} and Figure~\ref{fig:burkertfitsresiduals}). 

\begin{table}[H]
\caption{Diagnostics of the Burkert profile fits for Draco, Sextans, and UMi dwarfs.}
\label{tab:burkert_diagnostics}
\begin{tabularx}{\textwidth}{l CCC}
\toprule
\textbf{Diagnostic} & \textbf{Draco} & \textbf{Sextans} & \textbf{UMi} \\
\midrule
Mean acceptance fraction 
& 0.71
& 0.71 
& 0.71 \\
Reduced $\chi^2$ 
& 0.87 
& 0.70 
& 0.96 \\
\bottomrule
\end{tabularx}
\end{table}
\vspace{-8pt}

\section{~}
\textls[-30]{Here we show the figures related to the posterior distributions of the B+IMBH/NSC~models.}
\begin{figure}[H]
 \includegraphics[width=0.75\textwidth]{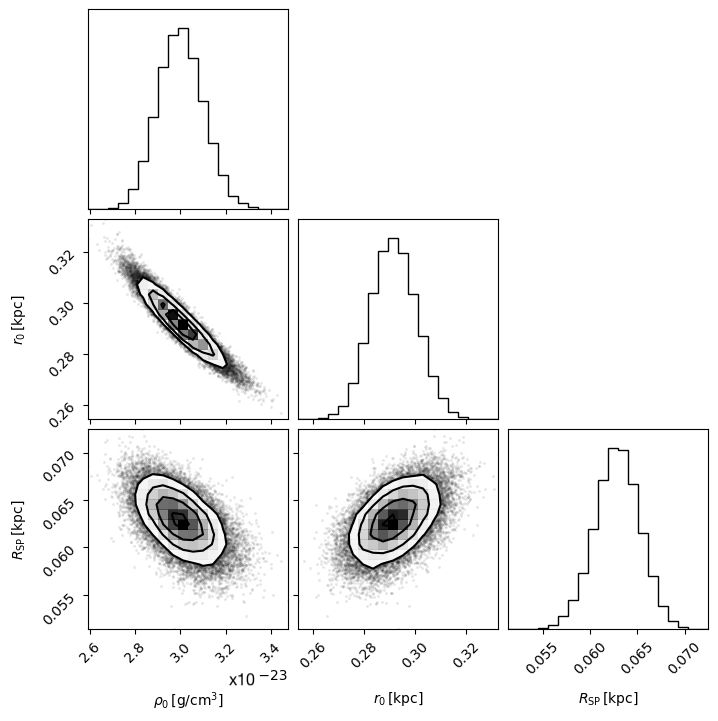}
 \caption{ {Draco} %
 posterior distributions of Burkert + IMBH spike fit for 
 $\rho_0$, $r_0$, and $R_{\mathrm{SP}}$ with their 1--3$\sigma$ credible contours 
 and median values.}
 \label{fig:corner_draco}
\end{figure}
\vspace{-6pt}

\begin{figure}[H]
 \includegraphics[width= 0.75\textwidth]{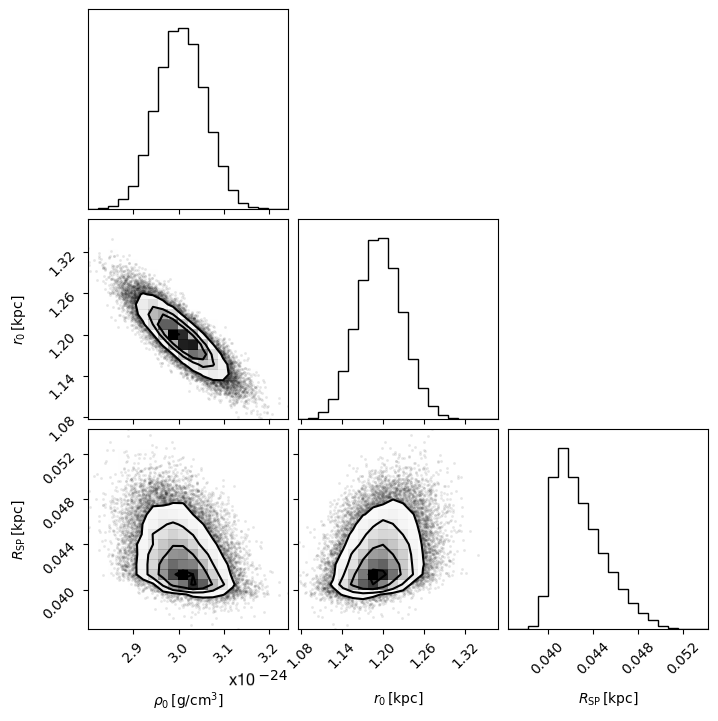}
 \caption{ {Sextans} %
 posterior distributions 
of Burkert + IMBH spike fit for $\rho_0$, $r_0$, and $R_{\mathrm{SP}}$ with their 1--3$\sigma$ credible 
 contours and median values.}
 \label{fig:corner_sextans}
\end{figure}
\vspace{-6pt}

\begin{figure}[H]
 \includegraphics[width= 0.75\textwidth]{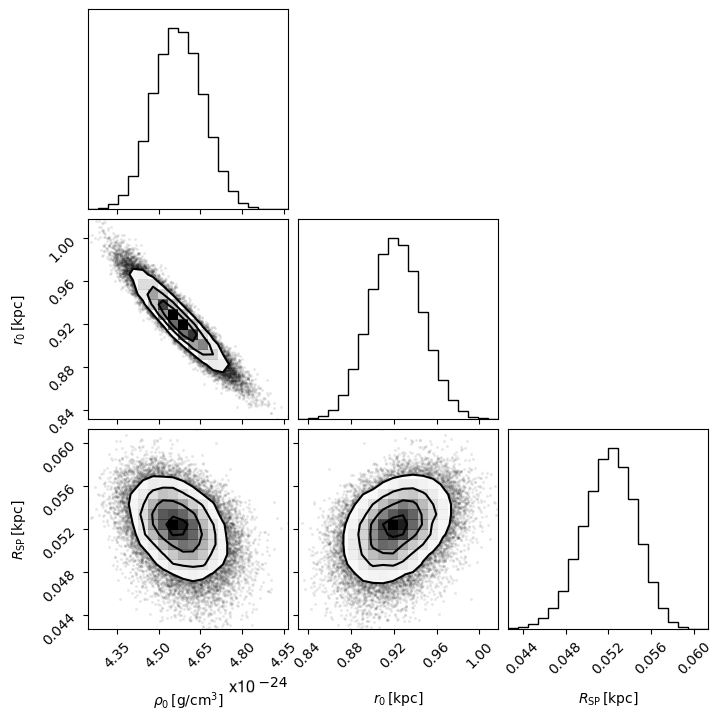}
 \caption{ {UMi} %
 posterior distributions of Burkert + IMBH spike fit for 
 $\rho_0$, $r_0$, and $R_{\mathrm{SP}}$ with their 1--3$\sigma$ credible contours 
 and median values.}
 \label{fig:corner_umi}
\end{figure}
\vspace{-6pt}

\begin{figure}[H]
 \includegraphics[width= 0.75\textwidth]{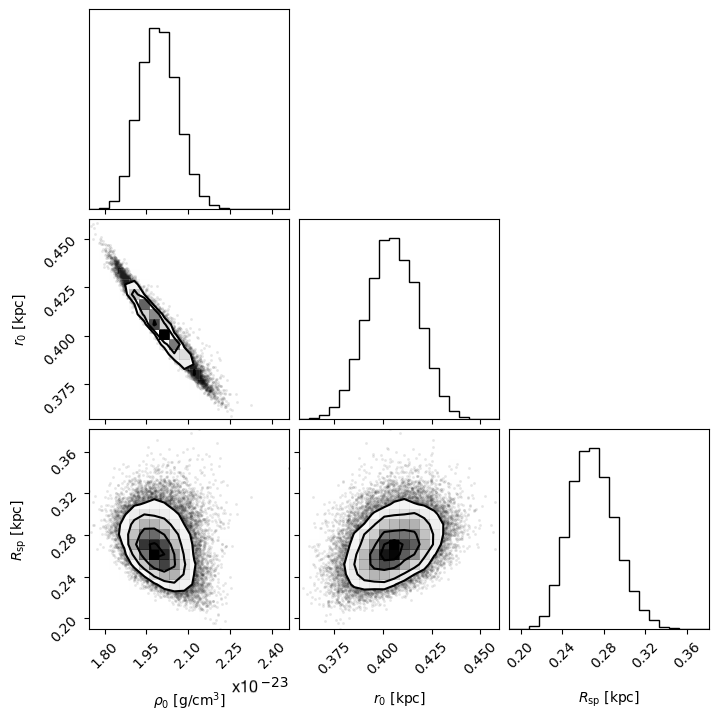}
 \caption{ {Draco} %
 posterior distributions of Burkert + NCS spike ($g=-1$) fit for 
 $\rho_0$, $r_0$, and $R_{\mathrm{SP}}$ with their 1--3$\sigma$ credible contours 
 and median values.}
 \label{fig:12}
\end{figure}
\vspace{-6pt}

\begin{figure}[H]
 \includegraphics[width= 0.75\textwidth]{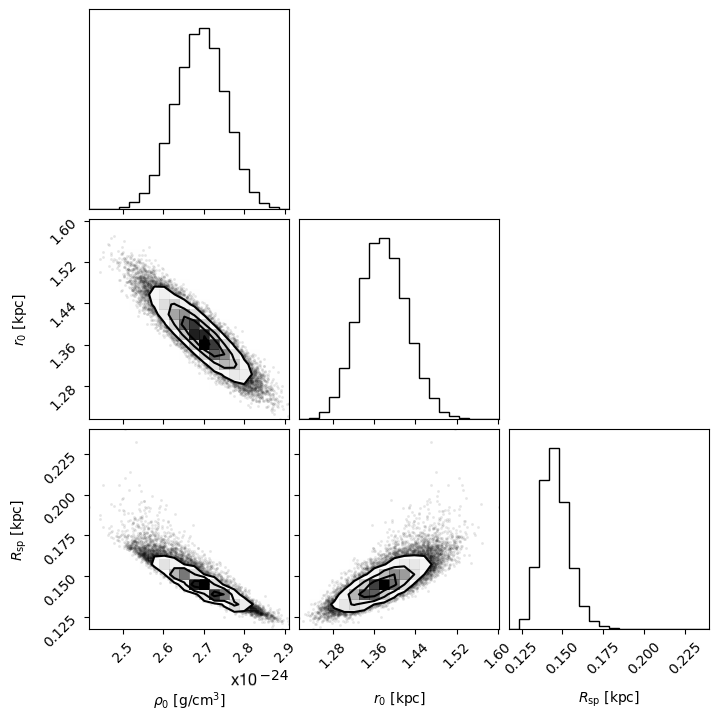}
 \caption{ {Sextans} %
 posterior distributions of Burkert + NCS spike ($g=-0.5$) fit for 
 $\rho_0$, $r_0$, and $R_{\mathrm{SP}}$ with their 1--3$\sigma$ credible contours 
 and median values.}
 \label{fig:13}
\end{figure}
\vspace{-6pt}

\begin{figure}[H]
 \includegraphics[width= 0.75\textwidth]{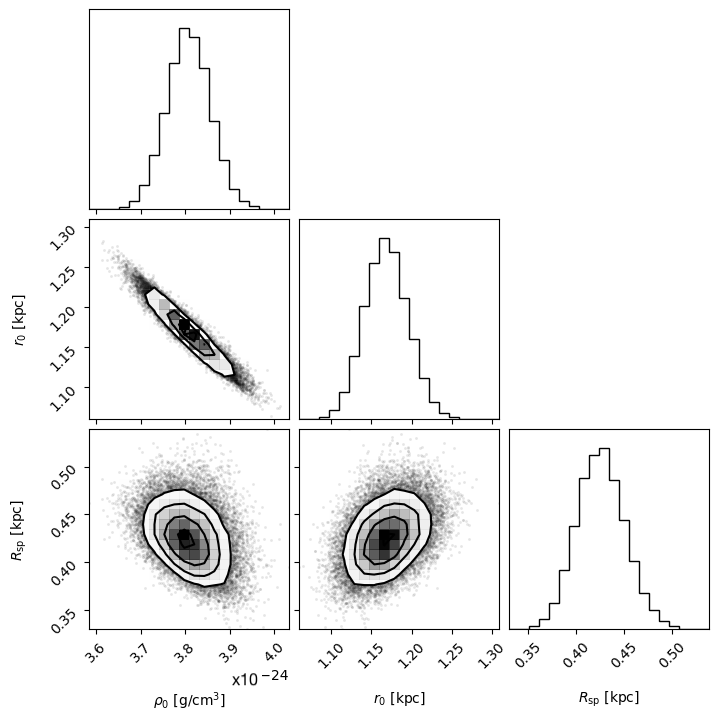}
 \caption{ {UMi} %
 posterior distributions of Burkert + NCS spike ($g=-0.5$) fit for 
 $\rho_0$, $r_0$, and $R_{\mathrm{SP}}$ with their 1--3$\sigma$ credible contours 
 and median values.}
 \label{fig:14}
\end{figure}

\begin{adjustwidth}{-\extralength}{0cm}
\printendnotes[custom] 

\reftitle{References}

 \PublishersNote{}
\end{adjustwidth}
 
\end{document}